\documentclass[superscriptaddress,reprint,longbibliography]{revtex4-1}

\usepackage{amssymb,amsmath}
\usepackage{graphicx,import}
\usepackage{placeins}
\usepackage{hyperref}
\usepackage{color}
\usepackage{longtable}
\usepackage{amsthm}
\usepackage{algorithm}
\usepackage{algpseudocode} 
\usepackage{upgreek}
\usepackage[nameinlink,poorman]{cleveref}

\graphicspath{{./paperfigs/}} 
\usepackage{enumitem}
\newlist{thmprop}{enumerate}{2}
\setlist[thmprop]{label={\normalfont(\roman*)},ref=(\roman*)}

\crefname{equation}{Eq.}{Eqs.}
\crefname{figure}{Fig.}{Figs.}
\crefname{table}{Table}{Tables} 
\crefname{section}{Section}{Sections}
\crefname{chapter}{Chapter}{Chapters}
\crefname{appendix}{Appendix}{Appendices}
\crefname{algorithm}{Algorithm}{Algorithms}

\crefname{theorem}{Theorem}{Theorems}
\crefname{defn}{Definiton}{Definitions}
\crefname{azm}{Assumption}{Assumptions}
\crefname{corollary}{Corollary}{Corollaries}
\crefname{lemma}{Lemma}{Lemmas}
\crefname{thmprop}{property}{properties}
\crefname{proposition}{Proposition}{Propositions}
\crefname{remark}{Remark}{Remarks}

\newcommand{\norm}[1]{\left\lVert#1\right\rVert}

\usepackage[normalem]{ulem}
\begin{document}

\title{Integration of spectator qubits into quantum computer architectures for hardware tuneup and calibration}

\author{Riddhi Swaroop Gupta} 
\email{riddhi.sw@gmail.com}
\affiliation{ARC Centre of Excellence for Engineered Quantum Systems, School of Physics, The University of Sydney, New South Wales 2006, Australia}

\author{Luke C.~G.~Govia}
\affiliation{Raytheon BBN Technologies, 10 Moulton St., Cambridge, MA 02138, USA}

\author{Michael J. Biercuk}
\affiliation{ARC Centre of Excellence for Engineered Quantum Systems, School of Physics, The University of Sydney, New South Wales 2006, Australia}

\begin{abstract} Performing efficient quantum computer tuneup and calibration is essential for growth in system complexity. In this work we explore the link between facilitating such capabilities and the underlying architecture of the physical hardware. We focus on the specific challenge of measuring (``mapping'') spatially inhomogeneous quasi-static calibration errors using spectator qubits dedicated to the task of sensing and calibration. We introduce a novel architectural concept for such spectator qubits: arranging them spatially according to prescriptions from optimal 2D approximation theory. We show that this insight allows for efficient reconstruction of inhomogeneities in qubit calibration, focusing on the specific example of frequency errors which may arise from fabrication variances or ambient magnetic fields. Our results demonstrate that optimal interpolation techniques display near optimal error-scaling in cases where the measured characteristic (here the qubit frequency) varies smoothly, and we probe the limits of these benefits as a function of measurement uncertainty. For more complex spatial variations, we demonstrate that the NMQA formalism for adaptive measurement and noise filtering outperforms optimal interpolation techniques in isolation, and crucially, can be combined with insights from optimal interpolation theory to produce a general purpose protocol. 
\end{abstract}

\maketitle

\section{Introduction} 

The scale-up of near-term quantum devices presents new challenges associated with fabrication, tuneup, characterization, and control as the total number of qubits increases  \cite{preskill2018quantum,kelly2018physical,tranter2018multiparameter,lennon2019efficiently,wigley2016fast}. For instance, in larger devices, we simultaneously see that fabrication tolerances may lead to performance variation across systems~\cite{Tolpygo_2015}, while the number of measurements required in order to bring systems online grows rapidly.  These concerns add to the general issue of decoherence due to undesired coupling of qubits to spatially inhomogeneous ambient fields. Accordingly, the procedure by which quantum devices must be characterized and tuned-up presents a challenge which grows with the size and complexity of each new generation of device~\cite{koch2020benchmarking, Arute_2019}.

In response, there has been an emergence of interest in the deployment of automated and nondestructive data inference and control approaches to this problem, exploiting the presence of spatial correlations in the performance variations~\cite{Arute_2019}. One interesting approach to this problem involves characterization of classical noise or device variations in space using spectator qubits \cite{majumder2020real}; additional dedicated devices are employed to gain information about hardware performance such that information-carrying data-qubits may operate undisturbed. Even within this framework we are still faced with the challenge of efficiently extracting information from potentially large numbers of these ancillary sensing devices. 

In such circumstances, classical adaptive filtering techniques can play an essential role in efficiently characterising, calibrating and controlling mesoscale quantum devices in an attempt to extract improved performance \cite{kelly2018physical,lennon2019efficiently,shankar2013autonomously}. In earlier work, the authors presented two distinct methods with radically different applications for the approximate reconstruction of unknown continuous physical phenomena using discretized projective measurements on qubits. The first method \cite{govia2018tomography} was an efficient approach to quantum state tomography of single-mode continuous variable systems using bivariate optimal Lagrange interpolation at the Padua points \cite{caliari2005bivariate,bos2006gencurve,bos2007ideal,caliari2011padua2dm,caliari2008bivariate}. The second method was inspired by probabilistic robotics and enabled adaptive scheduling of measurements on 2D multi-qubit arrays for noise spatial mapping, referred to as Noise-Mapping for Quantum Architectures (NMQA) \cite{gupta2019convergence,gupta2019autonomous}. 

In this work, we combine these apparently disparate techniques in order to address the problem of how to most efficiently extract useful information for mesoscale tuneup and calibration using spectator qubits. We model the underlying hardware performance variations as a ``field'' and thus seek the most efficient means to characterize the spatial field. To this end we are inspired by the observation that ideal interpolation theory provides insights on how to conduct optimal sampling on any continuously-varying, bivariate function (e.g. space, time, relative phase). We associate such optimal sampling points with the physical locations of sensor-qubits in order to efficiently determine spatial variations in the field at the location of the (unmeasured) data-qubits. 

In particular, we suggest sensor-qubits should be located on a grid defined by the so-called Padua points \cite{bos2006gencurve,bos2007ideal,caliari2011padua2dm,caliari2008bivariate}, which enables efficient Lagrange polynomial interpolation of the underlying field. We employ numeric simulations of field characterization using different algorithmic approaches to estimation and explore the influence of the underlying sensor-qubit locations on the quality of the estimates. We demonstrate that Lagrange-Padua reconstructions in 2D outperforms other interpolation or statistical estimation methods for the reconstruction of polynomial fields, while using $\sim10\times$ fewer sensor-qubits.  Our simulations show how these benefits are lost in conditions when optimal interpolation theory cannot be applied, and demonstrate that the alternative adaptive measurement approach NMQA~\cite{gupta2019autonomous,gupta2019convergence} achieves the lowest expected error. We study the impact of quantum projection noise on field estimation quality, including the role of averaging over the discretized outcomes of projective measurements. Our results demonstrate that NMQA is agnostic to the spatial arrangement of sensor-qubits on hardware, motivating the use of a Padua sensor grid as a default architectural choice for embedded sensors.

We outline our approach in \cref{sec:Approach}, where we present a classical spatial reconstruction and prediction problem over a 2D arrangement of sensor- and data-qubits. Two distinct types of device configurations are considered where a sensor-qubit sub-lattice is nested in a fixed data-qubit grid. For each device configuration, we introduce the appropriate spatial reconstruction strategy and test performance. Results from numeric simulations and comparative analyses are presented in \cref{sec:Results}.  We conclude with a summary and brief future outlook in \cref{sec:Conclusion}.

\section{Approaches to spatial field estimation via spectator qubits \label{sec:Approach}}

We consider a notional 2D multi-qubit array in which qubits are partitioned into `sensor-' and `data-' qubits; these non-interacting qubits are subject to spatial inhomogeneities in a device parameter ({\em e.g.} qubit frequency), modelled as an external classical field with finite spatial correlations. For concreteness we implement a term coupling to the qubit Hamiltonian $\propto\sigma_{z}$, as would be the case for qubit frequency variations or an inhomogeneous magnetic field. To probe this Hamiltonian term we assume a single-shot Ramsey experiment with a fixed interrogation time can be performed on each qubit, though the general framework is compatible with arbitrary measurement routines. The question arising in this work is the extent to which the spatial configuration of the sensor-qubits affects the accuracy of the field reconstruction at the proximal data-qubits. 

In the limit of high-fidelity measurements on each sensor-qubit, the task of mapping the field at the data-qubit locations transitions from an estimation problem to a bivariate interpolation or an approximation problem. Let $x$ (similarly, $x'$) correspond to the spatial coordinates of a sensor-qubit (data-qubit) in 2D, and let $f(x)$ (similarly, $f(x')$) be the estimated field value measured by repeated projective measurements. A key insight in this work is the recognition that some interpolation techniques are associated with an optimal choice of point-set at which to sample, $\chi:= \{x_i\}_{i=1}^{|\chi|}$, and that these points \emph{should} be associated with the spatial locations of sensor-qubits. The canonical example in 1D is the Chebshev points. In 2D, the Padua-points are a recently discovered family of point-sets for optimally sampling a bivariate continuous function, for which Lagrange polynomial interpolation yields the lowest-error reconstruction \cite{bos2006gencurve,bos2007ideal,caliari2011padua2dm,caliari2008bivariate}. 

The Padua points are families of point-sets that occur on the unit-square $[-1, 1]^2$. The geometric interpretation of Padua points can be obtained by envisioning them as equally spaced points along a generating curve on the unit square, defined by
\begin{align}
	\gamma_\kappa(t) := \left(- \cos((\kappa +1)t), -\cos(\kappa t)\right). \label{eqn:main:generatingcurve}
\end{align}
The intersection of this generating curve with itself, the edges of the square or its vertices yields the Padua point set. The Padua points are equispaced along $t$ and indexed by $j,j'$ as:
\begin{align}
	t_{(j,j')}&:= \frac{j\kappa + j' (\kappa +1)}{\kappa(\kappa + 1)} \pi, \quad j, j' \geq 0, j + j' \leq \kappa
\end{align} 
Each Padua point-set is associated with a positive integer \textit{order}, $\kappa$, which indicates that a Lagrange interpolant can be constructed at the Padua points using a family of polynomials with degree $ n \leq\kappa$. Different $\kappa$'s yield Padua point-sets, $\chi_\kappa$, that differ in both size and arrangement on the unit square, with the number of points in the set given by:
\begin{align}
 |\chi_\kappa| = \frac{(\kappa + 2)(\kappa + 1)}{2}. 
\end{align}
\noindent The structure of the Padua grid is unique with respect to the data-qubit grid for $\kappa>2$, while for $\kappa = 1, 2$, the Padua grid is degenerate with a square grid containing the relevant number of sites. 

For any point in the unit square $x' \in [-1,1]^2$, the Padua-Lagrange interpolant~\cite{bos2007ideal,bos2006gencurve} of a true field, denoted $f$, is written:
\begin{align}
	\mathcal{L} (f)(x') := \sum_{x \in \chi_\kappa} f(x) l(x, x').
\end{align}
\noindent Here the Lagrange basis polynomials, $l(x, x')$, form an orthonormal basis over the Padua points, such that for any two Padua points $x_1, x_2 \in \chi_\kappa$, $l(x_1, x_2) = 1 $ if and only if $ x_1=x_2$, and $l(x_1, x_2) = 0 $ otherwise \cite{bos2007ideal,caliari2011padua2dm}. If either one or both of $x_1, x_2$ is not a Padua point, $l(x_1, x_2) \geq 0$, and can be computed efficiently from the coordinate locations $x_1$ and $x_2$ using matrices involving the so-called `product-grids' of the Chebychev polynomials in 1D, as detailed in \cite{wigley2016fast}, and re-stated for completeness in the \textit{Appendices}. 

Locating sensor-qubits at Padua points may be contrasted with methods in which the sensor-qubit array is matched to the underlying geometry of the data-qubits, but incorporated as a nested, regular sub-lattice. In this circumstance, a polynomial basis need not be an optimal choice for interpolation. We may instead perform interpolation using radial basis functions (RBF) \cite{majdisova2017radial,stein2012interpolation}. The RBF method uses a non-polynomial (e.g.~sigmoidal or Gaussian) function basis to conduct interpolation, and scales well to multi-variate interpolation problems. 

So far we have discussed deterministic analysis methods such as interpolation, but we also consider nonlinear statistical estimation techniques that incorporate a projective-measurement model. We choose an algorithm presented in previous literature as NMQA~\cite{gupta2019convergence,gupta2019autonomous} which estimates an unknown 2D field by adaptively scheduling measurements on sensor-qubits. NMQA can be implemented directly on any sensor-qubit arrangement, including both regular (rectangular) lattices or Padua locations for sensor-qubits. However, not all of the available sensor-qubits need to be measured, as NMQA's controller selects the next physical measurement adaptively based on state estimation in the previous iteration step. It is thus distinct from interpolation methods as it focuses on building a ``map'' of $f$ iteratively as new measurements are obtained  using a spatial information sharing and prediction mechanism. NMQA procedures are characterized in detail in Refs.~\cite{gupta2019convergence,gupta2019autonomous}. 

\section{Results and Discussion \label{sec:Results}}

\begin{figure}[t!]
 \centering
 \includegraphics[scale=0.95]{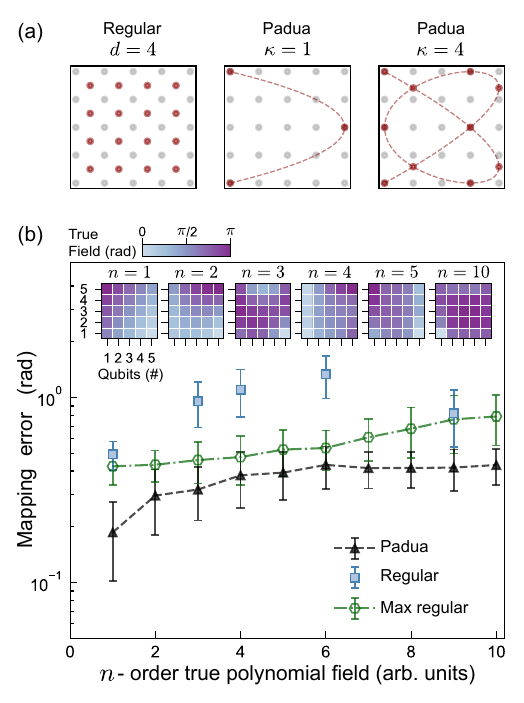}
 \caption[Impact of sensor-qubit layout on field mapping quality in a 2D device.]{  Impact of sensor-qubit layout on field mapping quality in a 2D device. (a) Fixed $5\times 5$ data-qubits (grey dots) used in all analyses. Left to right: sensor-qubits (red dots) are arranged in regularly-spaced $d\times d$ grid for $d=4$ (left) or according to Padua points of order $\kappa=1, 4$ (middle, right).  Generating curve for Padua point sets illustrated by red dashed line.  (b) Mapping error in radians (rad) vs. polynomial order $n$ for three different interpolation strategies. Top row of insets depicts an example true polynomial field of degree $n$ on a fixed $5\times 5$ regular grid of data-qubits.  Main panel: each data marker corresponds to an average over 50 trials of randomly generated true fields. In a single trial, a true field is generated as a polynomial of degree $n$ with random coefficients, and the output of the polynomial is scaled and shifted to lie between $[0, \pi]$ radians. Interpolation at Padua points of order $\kappa$ is performed using Lagrange polynomials with $\kappa=n$ (black dashed triangles). Interpolation on a $d\times d$ regular grid is performed using RBF where the total number of sensors is within $\pm 3$ qubits on both regular and Padua grids, with $d= 2, 3, 4, 5, 9$ (blue squares). For $d=9$, sensor-qubits that overlap with data-qubits are removed, leading to a total of $54$ sensor-qubits. In the extreme case where the number of sensor-qubits on regular grid are much greater than Padua grid for all orders $n<9$, RBF interpolation is performed on a fixed, regular grid using $d=9$ (green dashed markers). Error-bars represent 1 standard deviation within the distribution of 50 random trial polynomials. For all interpolation schemes $m=50$ measurements per sensor-qubit is used.}
 \label{fig:datafig:padvsrbf}
\end{figure}

In this section, we employ a numeric simulation to characterize the performance of these techniques for true fields exhibiting different forms of spatial variation, while incorporating different spatial arrangement of sensors on hardware. Our studies consider a fixed $5 \times 5 $ regular (square) grid of data-qubits with an overlaid grid of sensor-qubits arranged at locations dictated by the chosen estimation scheme (\cref{fig:datafig:padvsrbf}(a)).  Simulated measurements from the sensor-qubits are used to reconstruct a map of the field $f$ at the locations of the data-qubits on the array. For any mapping algorithm we calculate error via the uniform error, i.e.~the infinity norm, and refer to this quantity as the ``mapping error''.  The novel data inference procedures discussed here are compared against a baseline approach using a square grid of sensors and simply assigning sensor results to nearest-neighbor data-qubits (``Standard'' protocol).  Full details of all hardware geometries used in comparative simulations are provided in the \textit{Appendices}.

\subsection {Dependence of interpolation performance on sensor grid}

We begin by examining the dependence of mapping error on the choice of interpolation strategy and sensor grid in \cref{fig:datafig:padvsrbf}(b). Calculations take as input data the binary outcomes `0' or `1' of simulated projective single-qubit measurements and receive the same fixed total measurement budget.  Measurements are `batch-processed', and the total measurement budget is fixed as a multiple of the grid size, so that every sensor-qubit is measured $m$ times. Thus for $m>1$, measurements at every sensor-qubit are used to produce an estimate for the expectation value of the measurement, and this floating-point number is then given to the algorithm. 

The main panel in \cref{fig:datafig:padvsrbf}(b) shows simulated mapping error vs.~the polynomial order $n$ of different fields $f$, as achieved using different interpolation methods. Each data point in \cref{fig:datafig:padvsrbf}(b) shows the average mapping error and standard deviation for a set of $50$ randomly sampled polynomial fields of a given order $n$. Examples of true fields for different $n$ are depicted as a color-scale in the top row of insets. The mapping error for Lagrange-Padua interpolation is plotted for $\kappa=n$ and $m=50$ measurements per sensor-qubit (black dashed triangles). This dataset may be compared directly with a RBF interpolation using the maximum number of sensors possible irrespective of the order $n$ of the field $f$: a 54 sensor-qubit regular grid (green dashed) with an approximately $2:1$ ratio of sensor- to data- qubits.

Since the number of sensors in  Lagrange-Padua interpolation scales with $n$, we may also directly compare interpolation on a regular vs.~Padua grid with approximately the same number of sensor-qubits. The choices of $d\times d$ regular grid for $d=2,3,4,5,9$  and Padua grid with $\kappa = 1,3,4,6,9$ form different spatial geometries with approximately the same number of sensor-qubits. In these specific cases, we plot the output of RBF interpolation as a separate data-set (blue squares), where blue data-markers for polynomial order $n<5$ ($n\geq 5$) correspond to a maximum size difference of $\pm 1$ ($\pm 3$) sensor-qubits between regular vs. Padua grids.

In \cref{fig:datafig:padvsrbf}(b) we observe that for all polynomial orders $n$, the Lagrange-Padua interpolation approach outperforms both alternative RBF interpolation strategies, demonstrating the impact of the sensor-qubit architecture on interpolation quality. This observation holds irrespective of whether regular and Padua grids are approximately the same size or if RBF is provided a distinct advantage with a high number of sensor-qubits.  

The inclusion of error in sensor-qubit measurements due to quantum projection noise represents a deviation from the core assumptions of ideal interpolation theory (for both RBF and Padua), where the value of the field at the sensors is assumed to be known exactly. Indeed, comparisons to optimal interpolation theory are only formally facilitated in the regime where ideal measurements are obtained as $m\to\infty$.  Nonetheless, Lagrange-Padua reconstructions give low-error reconstructions of polynomial fields for $\kappa=n$ despite the fact that $m$ is finite, and that all interpolation methods receive noisy measurements at the location of sensor-qubits. 

\begin{figure*}[t!]
 \centering
 \includegraphics{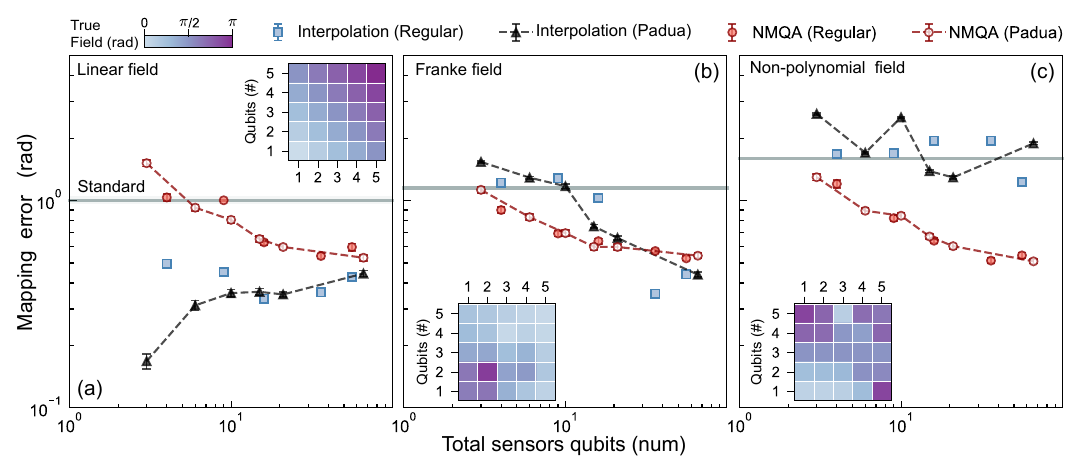}
 \caption[Performance of interpolation strategies and NMQA for reconstructing different true fields for both regular and Padua geometries]{Performance of interpolation strategies and NMQA for reconstructing different true fields for both regular and Padua geometries. True fields are depicted as a colorscale on a fixed $5\times5$ grid of data-qubits in insets. Main panels (a)-(c) show expected mapping error in radians (rad) vs.~number of sensors for (a) linear, (b) Franke function and (c) non-polynomial true field, with visible error bars depicting 1 std. error over $50$ repeated simulations. For Padua sensor locations, we plot NMQA (open red circles) vs. Lagrange polynomial interpolation (black dashed triangles). For a regular $d\times d$ grid, we plot NMQA (filled circles) vs. radial basis functions (blue squares). Standard assignment approach is shown as a threshold using a fixed arrangement of 9 sensor-qubits and averaged over 4 non-unique orientations (grey solid line). Data shown for $m=50$ measurements per sensor-qubit, with $d=2,3,4,6,9$ and $\kappa=1,2,3,4,5,10$.} 
 \label{fig:datafig:numofsensors}
\end{figure*}

\begin{figure}[t!]
 \includegraphics[scale=0.98]{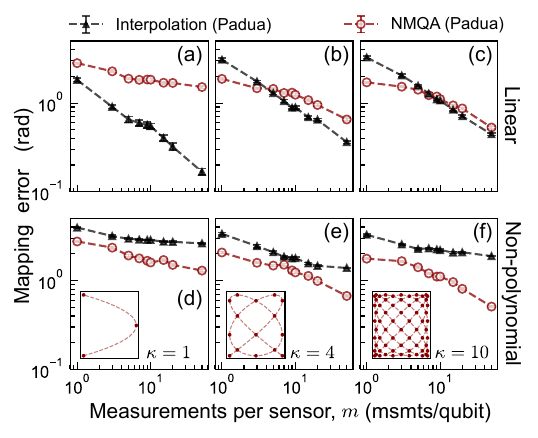}
 \caption[Expected mapping error vs. measurements per sensor-qubit $m$ for linear ($n=1$) and non-polynomial fields]{ Expected mapping error  in radians (rad) vs. measurements per sensor-qubit $m$ for linear field ($n=1$) in top row (a)-(c) and non-polynomial field in bottom row (d)-(f). Columns depict increasing order of the Padua grid from left to right $\kappa=1,4, 10$. In each panel, mapping error vs. $m$ number of measurements per sensor-qubit is plotted for NMQA (open red circles) vs. Lagrange polynomial interpolation (black dashed triangles) over $50$ trials with visible error bars depicting 1 std. error. Padua sensor arrangements (red dots) and associated generating curve (red dashed) for each column is shown as bottom left insets in (d)-(f). For a linear field in (a), $\kappa=n=1$ is optimal Lagrange-Padua interpolation for $m \to \infty$ and (b)-(c) represent an unnecessary increase in the number of sensor-qubits. In (d)-(f), a non-polynomial true field permits no finite order polynomial representation.}
 \label{fig:datafig:msmtpersensor}
\end{figure}

\subsection{Comparison between interpolation and adaptive measurement strategies}

Having confirmed the utility of the Lagrange-Padua approach, we now relax assumptions of ideal interpolation theory in \cref{fig:datafig:numofsensors}. We plot average mapping error vs.~the total number of sensor-qubits, but we change the true field so that a representation on the finite polynomial basis increases in difficulty from (a) to (c), which have fixed underlying fields. For each data point, we simulate 50 repetitions of the measurement outcomes for the same underlying field, and report an average mapping error and standard error.

Panel (a) in \cref{fig:datafig:numofsensors} corresponds to a simple linear variation, and in panel (b) we plot a common test-function in interpolation literature known as the Franke function \cite{franke1979critical}. This function consists of a weak linear background favouring a Lagrange-Padua approach, but with the addition of a superposition of two Gaussian functions, favouring the Gaussian basis in RBF interpolation. In panel (c), the true field is non-polynomial with a form $\propto \cos(\exp(2x+y))\sin(y)$ that does not permit any finite-polynomial or Gaussian representation.

The deviation from fields exactly describable by polynomial functions motivates the inclusion of the alternate adaptive mapping strategy, NMQA. In NMQA, a single-qubit binary measurement is fed iteratively into the algorithm; NMQA revises the state estimation procedure after each measurement, and the NMQA controller iteration adaptively chooses which sensor-qubit to measure next. This adaptive measurement procedure means that for any fixed total measurement budget, sensor-qubits under NMQA are not all uniformly measured $m$ times. 

We compare the different interpolation and adaptive measurement strategies across these three different fields using simulations setting $m=50$, but varying the number of sensor-qubits employed. \cref{fig:datafig:numofsensors}(a) numerically demonstrates that the Lagrange-Padua interpolation method for low order $\kappa$ performs the best of our chosen methods. This benefit arises because the underlying field is a degree-1 polynomial. In this case, the simple structure of the underlying field makes the use of additional sensor-qubits with $\kappa > 1$ unnecessary, as measurement errors from quantum projection noise for each additional sensor increase the total error in the reconstruction. This increase in reconstruction error is illustrated by the gradual increase in mapping error with sensor-qubit number for the black dashed triangles. The RBF interpolation strategy initially performs worse than Lagrange-Padua interpolation, but the two methods become approximately equivalent as more qubits are added, even though the field complexity remains fixed. In this case NMQA performs substantially worse than Lagrange-Padua interpolation for small qubit numbers, but its in-built noise filtering properties lead all methods to converge in the high $m, \kappa$ regime.

Moving to other functional forms for $f$, our results for the Franke function in \cref{fig:datafig:numofsensors}(b) show that the performance of the Lagrange-Padua method improves with increasing $\kappa$, but does not out-perform other methods, particularly for small or moderate values of $\kappa$, where it is not even well distinguished from the Standard assignment approach. RBF's superior performance in the high-data limit is somewhat unsurprising since Gaussian components of the Franke function are much stronger than the background linear drift, and these Gaussian spatial variations are naturally expressed in the functional basis of RBF. In this case, NMQA performs marginally better in the low-data regime but is outperformed by RBF in the large-measurement limit. 

Finally, we observe strikingly different behavior when $f$ is not approximated well by the natural functional basis of any interpolant. The final panel \cref{fig:datafig:numofsensors}(c) represents a challenging field reconstruction problem for all of our chosen methods. Interpolation using Lagrange polynomials on Padua grids vs.~RBF on regular grids shows that both interpolation methods perform similarly and provide little if any benefit relative to the Standard assignment approach. In this case we observe that NMQA substantially outperforms any other reconstruction strategy, illustrating the relevance of an adaptive measurement strategy when one cannot use an interpolant in some ideal functional basis relative to the true field. 

A surprising feature of panels \cref{fig:datafig:numofsensors}(a)-(c) is that the error scaling behaviour of NMQA appears to depend only on the total number of sensor-qubits on the unit square, and not on the spatial arrangement of sensor-qubits (Padua vs.~regular grid), unlike interpolation.   The difference between NMQA and interpolation strategies can also examined by varying the number of single-qubit measurements per sensor-qubit, $m$.   In  \cref{fig:datafig:msmtpersensor}, we plot mapping error vs.~$m$ for two different field configurations (rows) and three different orders for the sensor grid (columns). The limit $m\to\infty$ implies that a true field is known perfectly on the locations of the sensor-qubits for ideal interpolation, indicating that errors in functional evaluation are reduced from left to right along the $x$-axis in each panel of \cref{fig:datafig:msmtpersensor}. 

In \cref{fig:datafig:msmtpersensor}(a), $\kappa=n=1$ is close to satisfying conditions for optimal interpolation on a linear field and Lagrange-Padua method provides the lowest error reconstruction for all $m$. The opposite is true in (d)-(f), where NMQA significantly outperforms Lagrange-Padua on a non-polynomial field. For the panels with $\kappa > 1$ the performance of Lagrange-Padua methods deteriorates with respect to NMQA due to quantum projection noise. In some cases (e.g.~panels (b) and (c)), we see a crossover in performance between  Lagrange-Padua interpolation and NMQA with $m$ (as projection noise decreases with $m$). However, this behaviour disappears in the more challenging field reconstruction settings of (e) and (f), where NMQA's benefits are more substantial. A comparison of open circles in the top vs. bottom row of panels shows that mapping error trajectories for NMQA do not change with the type of true field. This observation empirically confirms that, unlike interpolation, NMQA's efficacy depends on branching random processes \cite{gupta2019convergence}, and hence its structure precludes choosing a functional basis for map reconstruction.

If \emph{a priori} information about the true field is available, the results of \cref{fig:datafig:numofsensors,fig:datafig:msmtpersensor} demonstrate that the appropriate configuration of Lagrange-Padua methods will yield the lowest error field reconstruction on unmeasured data-qubits. If no such information is available, then NMQA satisfies our physical intuition that reconstruction performance improves with an increase in the number of sensor-qubits, whereas this intuition is not always true for RBF or Lagrange interpolation if the interpolant is poorly specified with respect to the properties of the true field.

One anticipates that if $\kappa$ is sufficiently high relative to the polynomial order of the true field, and functional evaluation errors approach zero for large $m$, then Lagrange-Padua will always reflect a low-error reconstruction relative to RBF and NMQA approaches. Since there is no finite polynomial representation of the test-function in \cref{fig:datafig:numofsensors}(c), we have not been able to numerically ascertain the value of $\kappa$ that results in Lagrange-Padua interpolation outperforming NMQA.

\section{Conclusion \label{sec:Conclusion}}

Under the spectator qubit paradigm, we investigated protocols to efficiently characterize spatial inhomogeneities in qubit calibration or performance, modelled as a classical scalar field in 2D. By collecting measurements on a dedicated sub-lattice of sensor-qubits, we estimated the field values at the locations of proximal, unmeasured data-qubits in a 2D multi-qubit device, assuming a dephasing-type Hamiltonian. Drawing from optimal interpolation and statistical estimation theory, we used simulated data to compare the performance of reconstruction methods in the presence of different field configurations, studying the impact of the underlying sensor-qubit architecture of field characterization. 

Our results showed that in most circumstances it is advantageous to arrange sensor-qubits at the Padua points, an optimal point-set for 2D interpolation.  In circumstances where the field is well approximated by polynomial functions,  Lagrange-Padua interpolation outperforms comparable interpolation strategies using regular sensor-qubit lattices, as well as adaptive measurement strategies.  If, however, fields are not well approximated by polynomial functions, adaptive inference procedures such as NMQA perform best. Crucially, as we have shown here, the performance of NMQA shows limited sensitivity to the underlying sensor arrangement.

For complicated spatial variations, our results indicate that both finite-measurement effects and the presence of non-polynomial field structure reduce the utility of optimal Lagrange-Padua or RBF interpolation strategies in a spectator-qubit paradigm. In some cases \emph{a priori} information about spatial variation in a target field parameter may be available, but unless the interpolant can be appropriately specified, increasing the number of sensor-qubits can increase overall field-reconstruction error. This observation runs counter to the physical intuition that providing more sensors in a field reconstruction problem should improve the quality of the reconstruction. By contrast, the noise-filtering properties of adaptive strategies such as NMQA enable the algorithm to reduce mapping error as the total amount of information increases ({\em e.g.} increasing sensor counts). 

In this application NMQA also natively accommodates the potential to incorporate temporal drifts which are not compatible with interpolation strategies. The dynamical model in NMQA can be modified to include temporal dynamics that may either be specified \emph{a priori} or learned through data. In the adaptive control \cite{landau2011adaptive} and probabilistic robotics literature \cite{thrun2000probabilistic}, particularly in the context of spatio-temporal mapping applications for self-navigating vehicles \cite{wavesense}, a time-varying environment is often sampled rapidly such that a formal dynamical model need not be specified and estimated state information is `forgotten' by the algorithm over some time-scale. Contemporary inference proposals use hidden Markov auto-regressive moving average models \cite{michalek2000new} or Gaussian processes \cite{rasmussen2010gaussian} to introduce temporal correlations to an otherwise static pattern estimation problem \cite{luttinen2012efficient,hartikainen2011sparse,li2012arma}. A full spatio-temporal analysis of NMQA remains an exciting subject for future work. 

A hybrid of both NMQA and Lagrange-Padua techniques yield a general protocol for spatial field characterisation. For higher accuracy, one may first use NMQA for a coarse grained characterisation of a target field, followed by a Lagrange-Padua method on local regions that are approximately polynomial (with interpolation order and number of measurements informed by the results of NMQA). For quasi-static linear or quadratic spatial variation, only first or second order polynomial interpolants are needed, in which case the Padua point-set is an exact subset of a regular square grid, as explained in \textit{Appendices}. Thus, NMQA characterisation and local refinements via Lagrange-Padua interpolation are practically viable even for regular grids. Alternatively, building 2D qubit arrays in Padua arrangements is equally attractive, as Padua-grid-configured hardware will not impede any NMQA-based characterisation procedures. We look forward to greater exploration of how physical-layer quantum computer architectures may be impacted by the control and characterization strategies in use.

\section*{Acknowledgments}
This work partially supported by the ARC Centre of Excellence for Engineered Quantum Systems CE170100009, the US Army Research Office under Contract W911NF-12-R-0012, and a private grant from H. \& A. Harley.\\

\section*{Code and Data Availability}
Access to the code-base and data required to reproduce all figures is provided via http://github.com/qcl-sydney/nmqa.\\

\section*{Author Contributions}
R.S. Gupta performed all analysis, led technical efforts supporting the results presented, and co-wrote the manuscript. L.C.G.~Govia proposed the Padua points for spatial mapping, provided technical guidance on interpolation and error analysis, and contributed to writing the manuscript. M. J. Biercuk set the original research direction, provided guidance on analysis, and co-wrote the manuscript.

\clearpage
\appendix
\begin{figure*}[t!]
  \centering
  \includegraphics{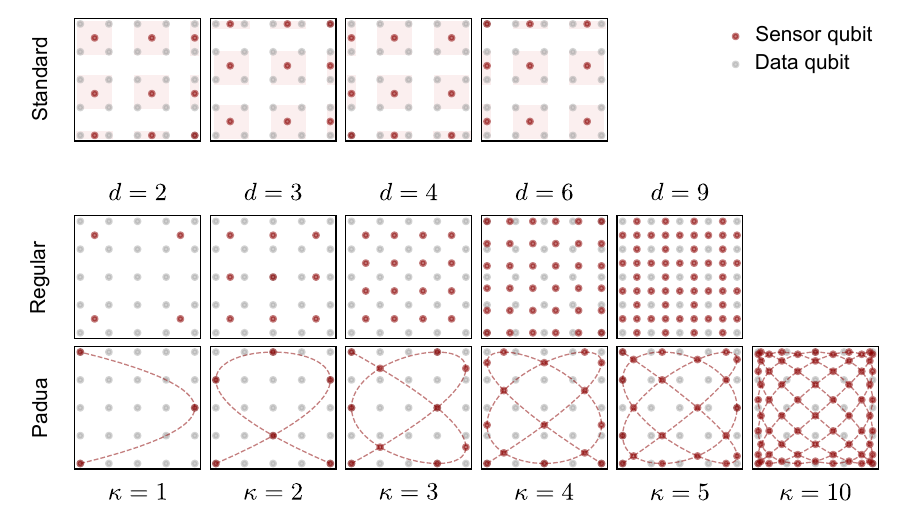}
  \caption[Placement of sensor and data-qubits on the unit square for Standard local neighborhood value-assignment (top), regular (middle) and Padua (bottom) configurations]{Placement of sensor and data-qubits on the unit square $[-1,1]^2$ for Standard local neighborhood value-assignment (top), regular (middle) and Padua (bottom) configurations. A $5\times5$ grid of data-qubits (grey circles) is fixed in all panels; while sensor-qubit (red circles) are varied. Top: four possible orientations of local neighbourhood assignment of data-qubits to a single sensor. Shaded regions depict fixed neighbourhood groups. Middle: regular nested grid of $d\times d$ sensors, with $d=2, 3, 4, 6$. For $d=9$, sensor-qubits that overlap exactly with data-qubits are removed, leading to a total of 54 sensor-qubits. Bottom: Padua locations for order $\kappa=1,2, 3, 4,5, 10$, and associated generating curve (red dashed). \label{fig:appendix:4}}
\end{figure*}

\section{Grid geometries for numerical simulation}
In \cref{fig:appendix:4} we illustrate the grid arrangements that have been used throughout the analysis of the main text. 

The top row of \cref{fig:appendix:4} shows four possible orientations for the ``Standard'' local neighbourhood assignment method that have been averaged over in all the data-figures presented. Local neighborhood assignments cannot generally be defined uniquely in 2D. For a regular data-qubit grid defined on a unit square, we choose the first four nearest-neighbour data-qubits for every sensor-qubit, thereby maximizing the number of first nearest neighbours in regular lattice. This procedure can continue at the interior points of any regular data-qubit grid when the number of rows and columns are even. For an odd row or column, the first nearest neighbours are necessarily equivalent to 1D case where two data-qubits are allocated to each sensor. The intersection of odd rows and columns is a vertex about which no neighbours are defined. We consider this asymmetric case for a $5\times5$ data-qubit grid and we average over the four possible orientations when we present results of data simulations.

The middle and bottom rows depict the regular and Padua spatial grid arrangements that have been investigated in the scope of our empirical study. Observing the Padua point-sets in the bottom row of \cref{fig:appendix:4}, we see that the Padua locations for $\kappa=1,2$ are an exact subset of a regular square grid arrangement. This suggests that quasi-static spatial variations for $\kappa \leq 2$ could be analyzed using Lagrange-Padua methods on specific types of regular grids.

\section{Ideal interpolation and approximation theory }

The details for the formalism of ideal interpolation and approximation theory is the subject of this section. We restate the theoretical formalism following \cite{van2015good} and establish notation to present the relevant results of \cite{ibrahimoglu2016lebesgue,bos2006gencurve,caliari2008bivariate,caliari2011padua2dm,bos2007ideal,kolomoitsev2018lp}.

\subsection{Least-squares criterion}

Let $\mathcal{C}(\Omega)$ be the space of continuous functions of two variables defined on an open, connected and bounded subset $\Omega \subset \mathbb{R}^2$ \cite{van2015good}. The geometry of $\Omega$ affects the interpolation problem and we focus on the case where $\Omega:=[-1, 1]^2$ takes the form of a unit square. Our true function is denoted by $f \in \mathcal{C}(\Omega)$ and $f$ is the subject of our approximation reconstruction over the region $\Omega$. 

We now wish to define two members in the space of bi-variate polynomials. Let $\mathcal{P}_\kappa$ be the space of bi-variate polynomials of degree at most $\kappa$ with dimension $N$. Let the points $\chi := \{ x_i \}_{i=1}^{L}$ denote the position of $L$ known values of $f$, and define a set of associated positive weights as $\mathcal{W} := \{ w_i \}_{i=1}^{L} \subset \mathbb{R}^+ \setminus \{0\}$. We define the best polynomial approximation to $f$ as $p^*$, and the minimum least squares polynomial as $p_L$ satisfying:

\begin{align}
 \min_{p_L} \parallel f - p_L \parallel_2 =\min_{p_L}
 \sqrt{\sum_{i=1}^L w_i^2 |f(x_i) - p_L(x_i)|^2}
\end{align}

This minimum least-squares operator is often re-stated in approximation problems as the interpolant $\mathcal{L}$ that maps $f$ to $p_L$, $\mathcal{L}: \mathcal{C}(\Omega) \to \mathcal{P}_\kappa$. The interpolant, $\mathcal{L}$, is thus a linear operator that depends on the point-set $\chi$, the weight-set $\mathcal{W}$ and the approximation space $\mathcal{P}_\kappa$ parameterized by the value $\kappa$.

The least squares approximating polynomial $p_L$ can be found by writing $\mathcal{L}$ in an appropriate polynomial basis, and solving for $p_L$. We recap the specific inverse problem and the error properties of $\mathcal{L}$ in terms of $\chi$ and $\kappa$ below.

Let $\{p_j\}_{j=1}^N$ be a polynomial basis for $\mathcal{P}_\kappa$ and $V_\chi:= [p_j(x_i)]$ be the $L \times N$ Vandermonde matrix for this basis using the point-set $\chi$. The interpolant at some arbitrary location $x \in \Omega$ is written in terms of coefficients in this basis:
\begin{align}
  \mathcal{L}(f)(x) &:= \rho(x)^T c = \sum_{j =1}^N p_j(x) c_j, \quad x \in \Omega \\
  c &:= \begin{bmatrix} c_1 & c_2 & \hdots & c_N \end{bmatrix}^T \\
  \rho &:= \begin{bmatrix} p_1 & p_2 & \hdots & p_N \end{bmatrix}^T
  \end{align} Here, $c$ represents $N$ scalar coefficients; $\rho$ represents an $N$-polynomial basis with each element corresponding to a basis-polynomial $p_j$. 
The coefficients $c$ are solutions to a linear system of equations $WV_\chi c = WF_\chi$, where:
\begin{align}
  W &: = \mathrm{diag}(w_1 \hdots w_L) \\
  F_\chi &: = \begin{bmatrix} f(x_1) & f(x_2) & \hdots f(x_L) \end{bmatrix}, \quad \{x_i\}_{i=1}^L \in \chi
\end{align} The coefficients are thus determined via optimisation or using the pseudo inverse $c= (WV_\chi)^\dagger W F_\chi $, yielding:
\begin{align}
  \mathcal{L}(f)(x) &:= \rho(x)^T (WV_\chi)^\dagger W F_\chi , \quad x \in \Omega
\end{align} Here, the matrix $W$ is invertible by construction. The pseudo-inverse governs error properties of the interpolant $\mathcal{L}$.

Following \cite{van2015good}, the statement that the solution $c$ is unique is equivalent to the condition that $V_\chi$ has full rank. If additionally $L=N$, then the point set $\chi$ is called uni-solvent for which $V_\chi$ is invertible and the interpolant simplifies to:
\begin{align}
  \mathcal{L}(f)(x) &:= \rho(x)^T V_\chi^{-1} F_\chi, \quad x \in \Omega
\end{align}

\subsection{Conditioning of the interpolant}
Using notation in the previous section, we now examine least squares solutions and add perturbations to the linear system of the previous section. In practical application of interpolation theory, these perturbations arise from inaccuracies in our choice of basis $V_\chi$ or the point-set $\chi$ that lead to inaccuracies in the solution $c$. For the purposes of this section, we present the concept of a condition number $\eta(V)$ for a choice of a basis for the matrix $V$. The condition number is unity for an optimal basis, which means that any errors in the matrix $V$ are not magnified as errors in the solution $c$. These errors are to be contrasted with the effect of measurement noise or errors in the data-set, which will be presented in future subsections, and thus, $F_\chi$ is noiseless in the discussion below.

Let the following equations represent a true linear system of equations and a perturbed linear system, denoted via ${}$ $\tilde{{}}$ ${}$ as:
\begin{align}
  WV_\chi c = WF_\chi \\ 
  W \tilde{V}_\chi \tilde{c} = WF_\chi, 
\end{align}
Next, we impose the $\infty$ or the uniform norm which satisfies consistency when applied to matrices and vectors such that $\norm{Ax} \leq \norm{A} \norm{x}$, where $A$ and $x$ represent arbitrary matrices and vectors. Let $E$ be an error matrix such that $\tilde{V}_\chi = V_\chi + E $
\begin{align}
   WF_\chi & = W (V_\chi + E) \tilde{c} = WV_\chi c \\
   \implies & (W V_\chi + W E) \tilde{c} = WV_\chi c \\
   \implies & (I + (WV_\chi)^\dagger W E) \tilde{c} = c\\
   \implies & \tilde{c} - c = (WV_\chi)^\dagger W E) \tilde{c}\\
   \implies & \norm{\tilde{c} - c} \leq \norm{(WV_\chi)^\dagger W E)} \norm{ \tilde{c}}
\end{align} In the above simplification, we assume $ (WV_\chi)^\dagger$ is a left pseudoinverse; and we apply consistency of the $\infty$-norm. If the quantity $ \norm{(WV_\chi)^\dagger W E)} < 1$, then the error in the estimated interpolation coefficients $\tilde{c}$ relative to the true (unknown) coefficients $c$ can be written as the following inequality \cite{stewart1996afternotes}:
\begin{align}
  \frac{\norm{\tilde{c} - c}}{\norm{c}} \leq \frac{\norm{(WV_\chi)^\dagger W E)}}{(1- \norm{(WV_\chi)^\dagger W E)})} 
\end{align} To interpret this inequality, we explicitly introduce the condition number by expanding and re-writing the term $\norm{(WV_\chi)^\dagger W E)}$ in terms of the condition number $\eta(V)$:
\begin{align}
\norm{(WV_\chi)^\dagger W E)} & \leq \norm{(WV_\chi)^\dagger}\norm{WE} \\
&= \norm{(WV_\chi)^\dagger}\norm{W\tilde{V}_\chi-WV_\chi }\\
&= \eta(V) \frac{\norm{W\tilde{V}_\chi-WV_\chi }}{\norm{WV_\chi}}\\
\eta(V)&:= \norm{(WV_\chi)^\dagger} \norm{WV_\chi}
\end{align} In this derivation, we apply the consistency of the uniform norm and we substitute the expression for $E$. The norm of the matrix $WV_\chi$ is added as a dummy variable, yielding the final expression for the condition number. When $\norm{(WV_\chi)^\dagger W E)} < 1$, we can write the following approximate inequality that links the condition number of $V_\chi$ as the magnification of true errors when any errors in the matrix $V_\chi$ are passed on to the solution for $c$:
\begin{align}
  \frac{\norm{\tilde{c} - c}}{\norm{c}} \leq \eta(V) \frac{\norm{W\tilde{V}_\chi-WV_\chi }}{\norm{WV_\chi}}\\
\end{align} It is well known that the optimal condition number is one, and in general, $\eta(V) \geq 1$ \cite{stewart1996afternotes}. 

The effect of the condition number on the $\infty$-norm of the interpolant can be seen by re-arranging terms to obtain $\norm{\mathcal{L}(f)} \leq \eta(V)\norm{\rho(x)^T} \norm{c}$. Note that the condition number does not depend on $f$ but only the structure of the linear system which we are trying to solve. Hence, for a general point set $\chi$ and choice of polynomial basis for $\mathcal{P}_\kappa$, $\eta(V)$ can be large.

Under an optimal choice of basis, however, the condition number satisfies $\eta(V)=1$ i.e. errors are not magnified in the inversion process \cite{van2015good}. This optimal basis for the least squares problem is in fact an ortho-normal polynomial basis with respect to a discrete inner product over $\chi$ \cite{van2015good}:
\begin{align}
  \langle p_i, p_j \rangle &:= \sum_{k=1}^L w_k^2 p_i(x_k) p_j(x_k)\\
  & = \delta_{i,j}, \quad \{x_k\}_{k=1}^L \in \chi, \{p_i\}_{i=1}^N
\end{align} 
The geometry of $\Omega$ influences whether an optimal basis can be found for a given application. 

\subsection{Lebesgue constant}

The $\infty$-norm of the interpolant $\norm{\mathcal{L}(f)}$ is typically used in approximation error analysis - in particular, to answer the question ``how good is $\mathcal{L}(f)$ as an approximation to $f$?'' To address this question, one seeks the so-called Lebesgue constant, $\Lambda_\mathcal{L}$, to provide the lowest upper bound on the norm of the interpolant:
\begin{align}
  \Lambda_\mathcal{L} := \min \{ \lambda \geq 0 \norm{\mathcal{L}(f)} \leq \lambda \norm{ f }, \quad \forall f \in \mathcal{C}(\Omega) \}
\end{align} Here, the constant $\Lambda_\mathcal{L}$ is independent of the form of the function $f$ or the construction of $\mathcal{L}(f)$. Hence, the behaviour of $\Lambda_\mathcal{L}$ enables a definition of optimal across different interpolation strategies or different geometries. 

In addition to providing a bound on the $\infty$-norm of the interpolant, we can also use $\Lambda_\mathcal{L}$ to provide an upper bound on approximation error. To do this, we consider three points in the space of continuous polynomials and apply the triangle inequality \cite{van2015good}. Since $\mathcal{P}_\kappa \subset \mathcal{C}(\Omega)$, we may consider $f, \mathcal{L}(f)$ and the best approximating polynomial $p^*$ as points in the space of continuous functions on $\Omega$. One applies the triangle inequality to obtain:
\begin{align}
  \parallel f - \mathcal{L}(f) \parallel \leq \parallel f - p^*\parallel + \parallel p^* - \mathcal{L}(f) \parallel, \quad \forall f \in \mathcal{C}(\Omega)
\end{align} Here, the best approximating polynomial is optimal with respect to the $\infty$-norm $\norm{f - p^*}$, and thus, it should be distingushed from the least-squares criterion used to construct the interpolant $\mathcal{L}(f)$.

Using $\mathcal{L}(p^*) := p^*$ and the linearity of the interpolant, one substitutes:
\begin{align}
\parallel p^* - \mathcal{L}(f) \parallel &= \parallel \mathcal{L}(p^* - f) \parallel \\ & \leq \Lambda_\mathcal{L} \parallel p^* - f \parallel \\
\implies \norm{ f - \mathcal{L}(f)} & \leq (1 + \Lambda_\mathcal{L})\norm{ f - p^*}
\end{align} The last step combines the two inequality relations and enables the interpretation of the Lebesgue constant as a measure of how much worse an interpolant performs with respect to some best approximating polynomial $p^*$.

\section{Sources of Error}
\subsection{Measurement errors in true $f$}
In this section, we analyse the effect of errors in the functional values at the point-set $\chi$. Let $\tilde{f}$ denote the perturbed function for the true $f $, with the error in functional values expressed as:
\begin{align}
  \epsilon(x) := f(x) - \tilde{f}(x), \quad f, \tilde{f} \in \mathcal{C}(\Omega); x \in \chi
\end{align}
The error in the interpolant is thus linear by the linearity of the operator $\mathcal{L}$:
\begin{align}
  \mathcal{L}(f) - \mathcal{L}(\tilde{f}) &:=\mathcal{L}(f - \tilde{f}) \\ & = \mathcal{L}(\epsilon)
\end{align} Letting $F_\chi$ and $\tilde{F}_\chi$ denote the corresponding vectors where each element is computed using $f$ and $\tilde{f}$ respectively, we obtain an expression for the error in the interpolant:
\begin{align}
  \mathcal{L}(\epsilon)(x) &:= \rho(x)^T (WV_\chi)^\dagger W (F_\chi - \tilde{F}_\chi), \quad x \in \Omega
\end{align}
If we assume further that $\epsilon \in \mathcal{C}(\Omega)$, we can use the definition of the Lebesgue constant:
\begin{align}
\norm{\mathcal{L}(\epsilon)} & \leq \Lambda_\mathcal{L} \norm{f - \tilde{f}}
\end{align}. 

Under the specific condition that $\norm{\mathcal{L}(f)} \geq \norm{f}$, the inequality above can be cast to establish $\norm{\mathcal{L}(f)}$ as the condition number for passing on errors in functional values to the interpolant:
\begin{align}
   \frac{\norm{\mathcal{L}(\epsilon)} }{\norm{\mathcal{L}(f)}}& \leq \Lambda_\mathcal{L} \frac{\norm{f - \tilde{f}}}{\norm{f}}
\end{align} In 1D, this condition is met for the Lagrange polynomials, leading to the straightforward interpretation that interpolation strategies with minimal $\Lambda_\mathcal{L}$ reduce sensitivity of the resulting interpolant to errors in functional values (see \cite{ibrahimoglu2016lebesgue} for 1D case). The condition $\norm{\mathcal{L}(f)} \geq \norm{f}$ is generally not true for an arbitrary polynomial basis for an interpolation strategy in 2D. For the specific case where $\chi$ represents the Padua point set on a unit square, one can derive the optimality relations discussed in subsequent sections. 

\subsection{Perturbed point-set $\chi$}
In this section, we assess the impact of a perturbed point-set $\chi$. For a chosen $\chi$ and polynomial basis for $\mathcal{P}_\kappa$, finite size effects mean that accessing exact $x \in \chi$ during practical applications is impossible.

There is very limited research on this subject (see \cite{austin2017trigonometric} for the case of 1D continuous function on an equispaced grid with a fixed perturbation); and as-yet no analysis for bi-variate interpolation problems on specific geometries (e.g. the unit square). 

In lieu of a formal derivation of error bounds, we consider the case that each point $x$ in the selected point-set $\chi$ is weakly perturbed by a fixed displacement $\epsilon \to 0$:
\begin{align}
  \epsilon := \tilde{x} - x, \quad \forall x \in \chi
\end{align}
The introduction of these errors now result in two non-linear perturbations: firstly, in the elements of $V_\chi \to V_{\chi, \epsilon}$ and similarly, $F_\chi \to F_{\chi, \epsilon}$, where the subscript ${}_{\chi, \epsilon}$ denotes that the matrix or vector elements are being computed using the perturbed points $\tilde{x}$.

For weak noise, the size of the perturbations depend on the derivatives of the function $f$ and the chosen polynomial basis $\{ p_j \}$ for the space $\mathcal{P}_\kappa$. 

A first order Taylor expansion recasts the effect of these perturbations on the functional values $F_{\chi, \epsilon}$ as approximately linear:
\begin{align}
  f(\tilde{x})&: = f(x) + \epsilon f'(x) + \mathcal{O}(\epsilon^2) \\
  \implies F_{\chi, \epsilon} &:= F_{\chi} + \epsilon F' + \mathcal{O}(\epsilon^2) \\
  F'&: = \begin{bmatrix} f'(x_1) & \hdots & f'(x_L) \end{bmatrix} 
\end{align} Thus, to first order, $F_{\chi, \epsilon}$ will manifest as measurement errors of the previous section. Higher orders may be considered depending on the specific form of $f$. The strength of the errors will depend on the derivatives of $f$. 

Similarly, first order expansions of the basis polynomials lead to an expression for the perturbed Vandermonde matrix:
\begin{align}
   p_j(\tilde{x}_i) &:= p_j(x_i) + \epsilon p_j'(x_i) + \mathcal{O}(\epsilon^2)\\
   \implies V_{\chi, \epsilon} &:= V_{\chi} + \epsilon V' + \mathcal{O}(\epsilon^2) \\
   V' &:= [p_j'(x_i)] \\
   j &= 1, \hdots, N \\
   i &= 1, \hdots, L 
\end{align} We now observe that the quantity $V'$ will have a first column consisting of all zeros; and the remaining elements form the basis for the polynomial space $\mathcal{P}_{\kappa -1}$. Hence $V'$ is singular, but a pseudo-inverse for the quantity $WV'$ exists.
We substitute these new matrices into the interpolant formula to get:
\begin{align}
  \mathcal{L}(\epsilon)(x) &:= \rho(x)^T (WV_{\chi} + \epsilon WV')^\dagger W (F_\chi + \epsilon F'). \quad x \in \Omega
\end{align}

We use the properties of the Moore-Penrose pseudoinverse for any pair of $m \times n$ matrices $A$ and $B$ \cite{tian2005moore}, 
\begin{align}
  (A + B)^\dagger = \frac{1}{2} \begin{bmatrix} I_n & I_n \end{bmatrix} 
  \begin{bmatrix} A & B \\ B & A \end{bmatrix}^\dagger \begin{bmatrix} I_m \\ I_m\end{bmatrix}
\end{align} to re-write the pseudo inverse of the first and second order $L \times N$ Vandermonde matrices:
\begin{align}
  \mathcal{V}(\epsilon) &:= (WV_{\chi} + \epsilon WV')^\dagger \\
  &= \frac{1}{2} \begin{bmatrix} I_N & I_N \end{bmatrix} 
  \begin{bmatrix} WV_{\chi} & \epsilon WV' \\ \epsilon WV' & WV_{\chi} \end{bmatrix}^\dagger \begin{bmatrix} I_L \\ I_L\end{bmatrix} \\
  &= \frac{1}{2\epsilon} \begin{bmatrix} I_N & I_N \end{bmatrix} 
  \begin{bmatrix} WV_{\chi}/\epsilon & WV' \\ WV' & WV_{\chi}/\epsilon \end{bmatrix}^\dagger \begin{bmatrix} I_L \\ I_L\end{bmatrix}, \epsilon \neq 0
\end{align} where the last line is obtained by using the basic property that the pseudoinverse of any non-zero scalar multiple of matrix $A$ is the reciprocal multiple of the pseudoinverse $A^\dagger$ satisfying $(\epsilon A)^\dagger = \epsilon^{-1} A^\dagger$, for some $\epsilon \neq 0$.

Under the weak perturbation approximation, and to first order in $\epsilon$, we see that the combined effect of a perturbed $\chi$ manifests as both errors in $\mathcal{V}(\epsilon)$ and in functional evaluation:
\begin{align}
  &\mathcal{L}(\epsilon)(x) = \nonumber \\ 
  & \quad \frac{1}{2}\rho(x)^T \begin{bmatrix} I_N & I_N \end{bmatrix} 
  \begin{bmatrix} WV_{\chi}/\epsilon & WV' \\ WV' & WV_{\chi}/\epsilon \end{bmatrix}^\dagger \begin{bmatrix} I_L \\ I_L\end{bmatrix}W ( \frac{F_\chi}{\epsilon} + F'),\\
  &\quad \epsilon \neq 0 \nonumber 
\end{align} In this form, it appears that the total error will be mediated by the condition number $\eta (\mathcal{V(\epsilon)})$ and additional magnification of order $\frac{1}{|\epsilon|}$ in functional evaluation in the weak error limit $|\epsilon| \to 0$. For $f$ approximately slowly-varying or constant, we can ignore errors in functional evaluation caused by the perturbed grid and set $F' \approx 0$, thereby noticing that an error $\epsilon$ inside the pseudo-inverse term and the functional evaluation term do not easily cancel out in general and only cancel if the off-diagonal first order terms additionally satisfy $V' \approx 0$. It remains an open question to see if a perturbed grid can be recast as functional evaluation errors under some other considerations. 

\section{Bivariate polynomial interpolation at the Padua points}

\subsection{Optimality on the unit square}

For the unit square, $\Omega := [-1, 1]^2$, it was discovered that the product Chebshev polynomials evaluated at the so-called Padua point-set $\chi_{p}$, and with the correct theoretically derived weights $W_{p}$ form an optimal basis with respect to the discrete inner product. Here, the subscript ${}_p$ denotes Padua-based interpolation strategies. Further, the interpolation problem presented in earlier sections turns out to be unisolvent, enabling a unique least squares solution for the interpolants $c_p$ and the matrix $W_pV_{\chi_p}$ turns out to have a unity condition number \cite{van2015good}. 

For the set of 2D problems mappable to the unit square, bi-variate Lagrange interpolation at the Padua points also yields the slowest-growing error bound for approximation error as $\Lambda_{\mathcal{L}_p} \sim \mathcal{O}(\log^2(\kappa))$ where $\kappa$ denotes the order of the Padua points to interpolate $f$ at most degree $\kappa$ \cite{caliari2011padua2dm,bos2007ideal,bos2006gencurve}. In particular, for some constant $a(f, k)$ that depends on $f \in \mathcal{C}^k(\Omega)$ and its $k$ continuous derivatives \cite{caliari2008bivariate}:
\begin{align}
  \norm{ f - \mathcal{L}(f)} & \leq (1 + \Lambda_\mathcal{L})\norm{ f - p^*} \leq a(f,k) \frac{\log^2(\kappa)} {\kappa^{k}} \label{eqn:error:lebesquebound}
\end{align} Here, $\kappa$ represents both the space of polynomials of degree at most $\kappa$ as well as the order of the Padua point set; $k$ denotes the number of continuous derivatives and $a$ is a constant that depends on both the function $f$ and $k$.

\subsection{Lagrange-Padua interpolation}
Padua points can be generated in 3 equivalent ways: (a) via the use of a generating curve approach \cite{bos2006gencurve}; the ideal theory approach \cite{bos2007ideal} and through the merger of two Chebychev-like grids \cite{caliari2011padua2dm}. There are four families of Padua points and we focus on the first family for the equations below.

Let $x \in \chi$ be a point in the set of Padua points $\chi$ of order $\kappa$ over the unit square $\Omega:=[-1, 1]^2$. The number of Padua points depends on the order, $\kappa$, as:
\begin{align}
	|\chi_\kappa|:= \frac{(\kappa+2)(\kappa+1)}{2}, \kappa> 0
\end{align}

The geometric interpretation of Padua points can be obtained by envisioning them as equally spaced points along a generating curve, $\gamma_\kappa(t)$ on the unit square, $\Omega$. The intersection of this generating curve with itself, the edges of the square or its vertices yields the Padua point set. The generating curve can be defined as:
\begin{align}
	\gamma_\kappa(t) := \left(- \cos((\kappa +1)t), -\cos(\kappa t)\right) \label{eqn:generatingcurve}
\end{align} On this curve, the Padua points are equispaced along $t$ and indexed by $j,m$ as:
\begin{align}
	t_{(j,m)}&:= \frac{j\kappa + m (\kappa +1)}{\kappa(\kappa + 1)} \pi, \quad j, m \geq 0, j+ m \leq \kappa
\end{align} The set of Padua points are classified as interior, boundary or vertex points. Two vertex points occur at $(1, 1)$ and $((-1)^\kappa, (-1)^{\kappa+1})$ and edge points occur on the boundary of the square. The curve of \cref{eqn:generatingcurve} is consistent with \cite{caliari2011padua2dm}.

The so called cubature weight $w_x$ of the point $x$ depends on its classification:
\begin{align}
	w_x &:= \frac{1}{\kappa (\kappa + 1)} \cdot
	\begin{cases} 
		1/2,&\quad x ~\mathrm{vertex} \\
		1, &\quad x ~\mathrm{boundary} \\
		2, &\quad x ~\mathrm{interior} \\
	\end{cases} 
\end{align} The cubature weights above agree with \cite{caliari2011padua2dm}. 

The formulae above enable interpolation for any true $f$ using Lagrange polynomial interpolation of degree at most $\kappa$. The interpolation formula is written as \cite{bos2007ideal,bos2006gencurve}:
\begin{align}
	\mathcal{L} (f)(x') := &\sum_{x \in \mathbb{P}_\kappa} f(x) w_x \left(K_\kappa(x, x') - T_\kappa(x[0]) T_\kappa(x'[0])\right) \\
	= & \sum_{x \in \mathbb{P}_\kappa} f(x) l(x, x') \\
	 l(x, x'):= &\frac{K^*(x,x')}{K^*(x,x)} \\
	K^*(x,y):= & K_\kappa(x,y) - T_\kappa(x[0]) T_\kappa(y[0]) \\
	K_\kappa(x,y):= & \sum_j^\kappa \sum_{i}^j T_i(x[0]) T_{j-i}(x[1]) T_i(y[0]) T_{j-i}(y[1]) \\
	w_x:= & \frac{1}{K^*(x,x)}
\end{align} where $(x[0], x[1])$ are the coordinates of $x$ in 2D, $T_\kappa(\cdot)$ are the Chebychev polynomials of order $\kappa$, $K_\kappa(x, y) $ is a reproducing kernel for the space of bivariate polynomials on the unit square with degree at most $\kappa$ \cite{bos2007ideal,caliari2011padua2dm}. The Lagrange basis polynomials form an orthonormal basis over the Padua points, having the property $l(x_1, x_2) = 1 \iff x_1=x_2$ and zero otherwise, for any two Padua points $x_1, x_2 \in \chi_\kappa$. 

We supplement the geometric picture of Padua points with an alternative formulation that enables rapid calculation. In this alternative picture, the Padua points are a subset of a grid of Chebyshev-Gauss-Lobatto points, $ \chi_\kappa \subset C_{\kappa+1} \times C_{\kappa+2}$. The interpolant can be written in terms of a matrix of coefficients , $\mathbb{C}_0(\cdot)$, and a rectangular Chebyshev matrix $\mathbb{T}(\cdot)$:
\begin{align}
	\mathcal{L} (f)(\textbf{X}):= &\left( (\mathbb{T}(X_1))^t \mathbb{C}_0(f) \mathbb{T}(X_2) \right)^t \\
	\mathbb{T}(S) := &
	\begin{bmatrix}
		\hat{T}_0(s_1) & \hdots & \hat{T}_0(s_m) \\
		\vdots & \hdots & \vdots \\
		\hat{T}_\kappa(s_1) & \hdots & \hat{T}_\kappa(s_m) 
	\end{bmatrix} \\
	S =& [s_1, \hdots, s_m], \quad \hat{T}_y(\cdot) := \sqrt{2} T_y(\cdot)
\end{align} Here, $\textbf{X} := X_1 \times X_2$ is a discrete Cartesian grid, $X_i$ is a vector of the $i$-th coordinate of all test points for the interpolation of the function $f$. The notation ${}^t$ denotes a matrix transpose. The Chebyshev matrix $\mathbb{T}(X_i)$ has dimensions $\kappa \times \mathrm{dims}(X_i)$ and has elements given by the scaled Chebyshev polynomials $\sqrt{2} T_y(\cdot)$ of order $y$.

The matrix of coefficients, $\mathbb{C}_0(f)$ is computed as essentially the left upper triangular component of a $(\kappa + 1) \times (\kappa + 1) $ square matrix $\mathbb{C}(f)$:
\begin{align}
 \mathbb{C}(f) := \mathbb{T}(C_{\kappa+1}) \mathbb{G}(f) (\mathbb{T}(C_{\kappa+2}))^t
\end{align} There is one modification: the last element of the first column of $\mathbb{C}_0(f)$ is multiplied by a factor of one-half. The construction of the remaining matrices will be discussed below. 

The rectangular Chebyshev matrices are now defined by a vectorised set of Chebyshev-Gauss-Lobatto points of order $\kappa + 1$:
\begin{align}
	C_{\kappa+1} := \{z_j^\kappa = -\cos((j-1)\pi/\kappa), \quad j=1, \hdots, \kappa+1\} \label{eqn:cgl:definition}
\end{align} In the above, the negative sign is required so that the Padua points, generating curve and the Chebyshev grids yield the same point-set.

Lastly, the $(\kappa + 1) \times (\kappa + 2) $ matrix $\mathbb{G}(f)$ incorporates the effect of the values of $f$ evaluated on Padua point set, as well as the Padua curbature weights. Entries of this matrix are non-zero only if the index of the matrix element coincides with a Padua point:

\begin{align}
	\mathbb{G}(f)&:= (g_{r,s}) \\ 
	&= 
	\begin{cases} 
		w_x f(x), &\quad \text{if $ x = (z_r^\kappa, z_s^{\kappa+1}) \in \chi_\kappa $} \\
		0, &\quad \text{if $ (z_r^\kappa, z_s^{\kappa+1}) \in C_{\kappa+1} \times C_{\kappa+2} \setminus \chi_\kappa $}
	\end{cases}
\end{align} The entries of $\mathbb{G}(f)$ which coincide with the Padua points can be quickly discovered by selecting only every other point in flattened `meshgrid' of $(\kappa + 1) \times (\kappa + 2) $ entries. The flattened mask is subsequently reshaped into a 2D matrix for odd values of $\kappa$ as $(\kappa + 1) \times (\kappa + 2) $. For even values of $\kappa$, the mask is reshaped to $(\kappa + 2) \times (\kappa + 1) $, followed by a matrix transpose. If this mask is applied to the meshgrid of the Chebyshev-Gauss-Lobatto points $C_{\kappa +1}\times C_{\kappa +2}$, the Padua points of the generating curve approach in the previous section are recovered. Computationally, it is easier to construct $\mathbb{G}(f)$ by generating and merging the two Chebyshev grids, $C_{\kappa+1}, C_{\kappa+2}$ to obtain the Padua points. 

The connection with the geometric interpretation is easier to see if we re-write the Padua points with slightly modified index notation \cite{bos2007ideal}:
\begin{align}
	x[0] & := \cos \frac{k \pi}{\kappa}, \quad 0 \leq k \leq \kappa \\
	x[1] & := 
	\begin{cases} 
		\cos \frac{(2j-1) \pi}{(\kappa + 1)}, \quad k ~\mathrm{even} \\
		\cos \frac{(2j-2) \pi}{(\kappa + 1)}, \quad k ~\mathrm{odd} \\
	\end{cases}
	& j = 1, \hdots \lceil \kappa/2\rceil + 1
\end{align} The formulae above result in duplicate Padua points that can be removed by inspection; direct comparison with the definition of the   Chebyshev-Gauss-Lobatto points confirms that $\chi_\kappa \subset C_{\kappa+1} \times C_{\kappa+2}$.

Two additional modifications to our Python code-base are required: firstly, the $\mathcal{L} (f)(\textbf{X})$ matrix is flipped from left-to-right (corresponding to the use of the left upper triangular matrix of coefficients). Secondly, $\mathcal{L} (f)(\textbf{X})$ is globally divided by a factor of four to compensate for the scaling factors in the rectangular Chebyshev matrix $\mathbb{T}(\cdot)$. Both of these modifications yield the final result in the correct orientation consistent with \cite{caliari2011padua2dm}.

\clearpage


\begin{thebibliography}{35}%
	\makeatletter
	\providecommand \@ifxundefined [1]{%
		\@ifx{#1\undefined}
	}%
	\providecommand \@ifnum [1]{%
		\ifnum #1\expandafter \@firstoftwo
		\else \expandafter \@secondoftwo
		\fi
	}%
	\providecommand \@ifx [1]{%
		\ifx #1\expandafter \@firstoftwo
		\else \expandafter \@secondoftwo
		\fi
	}%
	\providecommand \natexlab [1]{#1}%
	\providecommand \enquote  [1]{``#1''}%
	\providecommand \bibnamefont  [1]{#1}%
	\providecommand \bibfnamefont [1]{#1}%
	\providecommand \citenamefont [1]{#1}%
	\providecommand \href@noop [0]{\@secondoftwo}%
	\providecommand \href [0]{\begingroup \@sanitize@url \@href}%
	\providecommand \@href[1]{\@@startlink{#1}\@@href}%
	\providecommand \@@href[1]{\endgroup#1\@@endlink}%
	\providecommand \@sanitize@url [0]{\catcode `\\12\catcode `\$12\catcode
		`\&12\catcode `\#12\catcode `\^12\catcode `\_12\catcode `\%12\relax}%
	\providecommand \@@startlink[1]{}%
	\providecommand \@@endlink[0]{}%
	\providecommand \url  [0]{\begingroup\@sanitize@url \@url }%
	\providecommand \@url [1]{\endgroup\@href {#1}{\urlprefix }}%
	\providecommand \urlprefix  [0]{URL }%
	\providecommand \Eprint [0]{\href }%
	\providecommand \doibase [0]{http://dx.doi.org/}%
	\providecommand \selectlanguage [0]{\@gobble}%
	\providecommand \bibinfo  [0]{\@secondoftwo}%
	\providecommand \bibfield  [0]{\@secondoftwo}%
	\providecommand \translation [1]{[#1]}%
	\providecommand \BibitemOpen [0]{}%
	\providecommand \bibitemStop [0]{}%
	\providecommand \bibitemNoStop [0]{.\EOS\space}%
	\providecommand \EOS [0]{\spacefactor3000\relax}%
	\providecommand \BibitemShut  [1]{\csname bibitem#1\endcsname}%
	\let\auto@bib@innerbib\@empty
	\bibitem [{\citenamefont {Preskill}(2018)}]{preskill2018quantum}%
	\BibitemOpen
	\bibfield  {author} {\bibinfo {author} {\bibfnamefont {John}\ \bibnamefont
			{Preskill}},\ }\bibfield  {title} {\enquote {\bibinfo {title} {Quantum
				computing in the {NISQ} era and beyond},}\ }\href@noop {} {\bibfield
		{journal} {\bibinfo  {journal} {Quantum}\ }\textbf {\bibinfo {volume} {2}},\
		\bibinfo {pages} {79} (\bibinfo {year} {2018})}\BibitemShut {NoStop}%
	\bibitem [{\citenamefont {Kelly}\ \emph {et~al.}(2018)\citenamefont {Kelly},
		\citenamefont {O'Malley}, \citenamefont {Neeley}, \citenamefont {Neven},\
		and\ \citenamefont {Martinis}}]{kelly2018physical}%
	\BibitemOpen
	\bibfield  {author} {\bibinfo {author} {\bibfnamefont {Julian}\ \bibnamefont
			{Kelly}}, \bibinfo {author} {\bibfnamefont {Peter}\ \bibnamefont {O'Malley}},
		\bibinfo {author} {\bibfnamefont {Matthew}\ \bibnamefont {Neeley}}, \bibinfo
		{author} {\bibfnamefont {Hartmut}\ \bibnamefont {Neven}}, \ and\ \bibinfo
		{author} {\bibfnamefont {John~M}\ \bibnamefont {Martinis}},\ }\bibfield
	{title} {\enquote {\bibinfo {title} {Physical qubit calibration on a directed
				acyclic graph},}\ }\href@noop {} {\bibfield  {journal} {\bibinfo  {journal}
			{arXiv preprint arXiv:1803.03226}\ } (\bibinfo {year} {2018})}\BibitemShut
	{NoStop}%
	\bibitem [{\citenamefont {Tranter}\ \emph {et~al.}(2018)\citenamefont
		{Tranter}, \citenamefont {Slatyer}, \citenamefont {Hush}, \citenamefont
		{Leung}, \citenamefont {Everett}, \citenamefont {Paul}, \citenamefont
		{Vernaz-Gris}, \citenamefont {Lam}, \citenamefont {Buchler},\ and\
		\citenamefont {Campbell}}]{tranter2018multiparameter}%
	\BibitemOpen
	\bibfield  {author} {\bibinfo {author} {\bibfnamefont {Aaron~D}\ \bibnamefont
			{Tranter}}, \bibinfo {author} {\bibfnamefont {Harry~J}\ \bibnamefont
			{Slatyer}}, \bibinfo {author} {\bibfnamefont {Michael~R}\ \bibnamefont
			{Hush}}, \bibinfo {author} {\bibfnamefont {Anthony~C}\ \bibnamefont {Leung}},
		\bibinfo {author} {\bibfnamefont {Jesse~L}\ \bibnamefont {Everett}}, \bibinfo
		{author} {\bibfnamefont {Karun~V}\ \bibnamefont {Paul}}, \bibinfo {author}
		{\bibfnamefont {Pierre}\ \bibnamefont {Vernaz-Gris}}, \bibinfo {author}
		{\bibfnamefont {Ping~Koy}\ \bibnamefont {Lam}}, \bibinfo {author}
		{\bibfnamefont {Ben~C}\ \bibnamefont {Buchler}}, \ and\ \bibinfo {author}
		{\bibfnamefont {Geoff~T}\ \bibnamefont {Campbell}},\ }\bibfield  {title}
	{\enquote {\bibinfo {title} {Multiparameter optimisation of a magneto-optical
				trap using deep learning},}\ }\href@noop {} {\bibfield  {journal} {\bibinfo
			{journal} {Nature communications}\ }\textbf {\bibinfo {volume} {9}},\
		\bibinfo {pages} {1--8} (\bibinfo {year} {2018})}\BibitemShut {NoStop}%
	\bibitem [{\citenamefont {Lennon}\ \emph {et~al.}(2019)\citenamefont {Lennon},
		\citenamefont {Moon}, \citenamefont {Camenzind}, \citenamefont {Yu},
		\citenamefont {Zumb{\"u}hl}, \citenamefont {Briggs}, \citenamefont {Osborne},
		\citenamefont {Laird},\ and\ \citenamefont {Ares}}]{lennon2019efficiently}%
	\BibitemOpen
	\bibfield  {author} {\bibinfo {author} {\bibfnamefont {DT}~\bibnamefont
			{Lennon}}, \bibinfo {author} {\bibfnamefont {H}~\bibnamefont {Moon}},
		\bibinfo {author} {\bibfnamefont {LC}~\bibnamefont {Camenzind}}, \bibinfo
		{author} {\bibfnamefont {Liuqi}\ \bibnamefont {Yu}}, \bibinfo {author}
		{\bibfnamefont {DM}~\bibnamefont {Zumb{\"u}hl}}, \bibinfo {author}
		{\bibfnamefont {GAD}\ \bibnamefont {Briggs}}, \bibinfo {author}
		{\bibfnamefont {MA}~\bibnamefont {Osborne}}, \bibinfo {author} {\bibfnamefont
			{EA}~\bibnamefont {Laird}}, \ and\ \bibinfo {author} {\bibfnamefont
			{N}~\bibnamefont {Ares}},\ }\bibfield  {title} {\enquote {\bibinfo {title}
			{Efficiently measuring a quantum device using machine learning},}\
	}\href@noop {} {\bibfield  {journal} {\bibinfo  {journal} {npj Quantum
			Information}\ }\textbf {\bibinfo {volume} {5}},\ \bibinfo {pages} {1--8}
	(\bibinfo {year} {2019})}\BibitemShut {NoStop}%
\bibitem [{\citenamefont {Wigley}\ \emph {et~al.}(2016)\citenamefont {Wigley},
	\citenamefont {Everitt}, \citenamefont {van~den Hengel}, \citenamefont
	{Bastian}, \citenamefont {Sooriyabandara}, \citenamefont {McDonald},
	\citenamefont {Hardman}, \citenamefont {Quinlivan}, \citenamefont {Manju},
	\citenamefont {Kuhn} \emph {et~al.}}]{wigley2016fast}%
\BibitemOpen
\bibfield  {author} {\bibinfo {author} {\bibfnamefont {Paul~B}\ \bibnamefont
		{Wigley}}, \bibinfo {author} {\bibfnamefont {Patrick~J}\ \bibnamefont
		{Everitt}}, \bibinfo {author} {\bibfnamefont {Anton}\ \bibnamefont {van~den
			Hengel}}, \bibinfo {author} {\bibfnamefont {John~W}\ \bibnamefont {Bastian}},
	\bibinfo {author} {\bibfnamefont {Mahasen~A}\ \bibnamefont {Sooriyabandara}},
	\bibinfo {author} {\bibfnamefont {Gordon~D}\ \bibnamefont {McDonald}},
	\bibinfo {author} {\bibfnamefont {Kyle~S}\ \bibnamefont {Hardman}}, \bibinfo
	{author} {\bibfnamefont {Ciaron~D}\ \bibnamefont {Quinlivan}}, \bibinfo
	{author} {\bibfnamefont {P}~\bibnamefont {Manju}}, \bibinfo {author}
	{\bibfnamefont {Carlos~CN}\ \bibnamefont {Kuhn}},  \emph {et~al.},\
}\bibfield  {title} {\enquote {\bibinfo {title} {Fast machine-learning online
		optimization of ultra-cold-atom experiments},}\ }\href@noop {} {\bibfield
{journal} {\bibinfo  {journal} {Scientific reports}\ }\textbf {\bibinfo
	{volume} {6}},\ \bibinfo {pages} {1--6} (\bibinfo {year} {2016})}\BibitemShut
{NoStop}%
\bibitem [{\citenamefont {Tolpygo}\ \emph {et~al.}(2015)\citenamefont
	{Tolpygo}, \citenamefont {Bolkhovsky}, \citenamefont {Weir}, \citenamefont
	{Johnson}, \citenamefont {Gouker},\ and\ \citenamefont
	{Oliver}}]{Tolpygo_2015}%
\BibitemOpen
\bibfield  {author} {\bibinfo {author} {\bibfnamefont {Sergey~K.}\
		\bibnamefont {Tolpygo}}, \bibinfo {author} {\bibfnamefont {Vladimir}\
		\bibnamefont {Bolkhovsky}}, \bibinfo {author} {\bibfnamefont {Terence~J.}\
		\bibnamefont {Weir}}, \bibinfo {author} {\bibfnamefont {Leonard~M.}\
		\bibnamefont {Johnson}}, \bibinfo {author} {\bibfnamefont {Mark~A.}\
		\bibnamefont {Gouker}}, \ and\ \bibinfo {author} {\bibfnamefont {William~D.}\
		\bibnamefont {Oliver}},\ }\bibfield  {title} {\enquote {\bibinfo {title}
		{Fabrication process and properties of fully-planarized deep-submicron
			{Nb/Al}– $\hbox{AlO}_{\rm x}\hbox{/Nb} $ {Josephson} junctions for vlsi
			circuits},}\ }\href {\doibase 10.1109/tasc.2014.2374836} {\bibfield
	{journal} {\bibinfo  {journal} {IEEE Transactions on Applied
			Superconductivity}\ }\textbf {\bibinfo {volume} {25}},\ \bibinfo {pages}
	{1–12} (\bibinfo {year} {2015})}\BibitemShut {NoStop}%
\bibitem [{\citenamefont {Koch}\ \emph {et~al.}(2020)\citenamefont {Koch},
	\citenamefont {Martin}, \citenamefont {Patel}, \citenamefont {Wessing},\ and\
	\citenamefont {Alsing}}]{koch2020benchmarking}%
\BibitemOpen
\bibfield  {author} {\bibinfo {author} {\bibfnamefont {Daniel}\ \bibnamefont
		{Koch}}, \bibinfo {author} {\bibfnamefont {Brett}\ \bibnamefont {Martin}},
	\bibinfo {author} {\bibfnamefont {Saahil}\ \bibnamefont {Patel}}, \bibinfo
	{author} {\bibfnamefont {Laura}\ \bibnamefont {Wessing}}, \ and\ \bibinfo
	{author} {\bibfnamefont {Paul~M.}\ \bibnamefont {Alsing}},\ }\href@noop {}
{\enquote {\bibinfo {title} {Benchmarking qubit quality and critical
			subroutines on {IBM}'s 20 qubit device},}\ } (\bibinfo {year} {2020}),\
\Eprint {http://arxiv.org/abs/2003.01009} {arXiv:2003.01009 [quant-ph]}
\BibitemShut {NoStop}%
\bibitem [{\citenamefont {Arute}\ \emph {et~al.}(2019)\citenamefont {Arute},
	\citenamefont {Arya}, \citenamefont {Babbush}, \citenamefont {Bacon},
	\citenamefont {Bardin}, \citenamefont {Barends}, \citenamefont {Biswas},
	\citenamefont {Boixo}, \citenamefont {Brandao}, \citenamefont {Buell},\ and\
	\citenamefont {et~al.}}]{Arute_2019}%
\BibitemOpen
\bibfield  {author} {\bibinfo {author} {\bibfnamefont {Frank}\ \bibnamefont
		{Arute}}, \bibinfo {author} {\bibfnamefont {Kunal}\ \bibnamefont {Arya}},
	\bibinfo {author} {\bibfnamefont {Ryan}\ \bibnamefont {Babbush}}, \bibinfo
	{author} {\bibfnamefont {Dave}\ \bibnamefont {Bacon}}, \bibinfo {author}
	{\bibfnamefont {Joseph~C.}\ \bibnamefont {Bardin}}, \bibinfo {author}
	{\bibfnamefont {Rami}\ \bibnamefont {Barends}}, \bibinfo {author}
	{\bibfnamefont {Rupak}\ \bibnamefont {Biswas}}, \bibinfo {author}
	{\bibfnamefont {Sergio}\ \bibnamefont {Boixo}}, \bibinfo {author}
	{\bibfnamefont {Fernando G. S.~L.}\ \bibnamefont {Brandao}}, \bibinfo
	{author} {\bibfnamefont {David~A.}\ \bibnamefont {Buell}}, \ and\ \bibinfo
	{author} {\bibnamefont {et~al.}},\ }\bibfield  {title} {\enquote {\bibinfo
		{title} {Quantum supremacy using a programmable superconducting processor},}\
}\href {\doibase 10.1038/s41586-019-1666-5} {\bibfield  {journal} {\bibinfo
	{journal} {Nature}\ }\textbf {\bibinfo {volume} {574}},\ \bibinfo {pages}
{505–510} (\bibinfo {year} {2019})}\BibitemShut {NoStop}%
\bibitem [{\citenamefont {Majumder}\ \emph {et~al.}(2020)\citenamefont
	{Majumder}, \citenamefont {de~Castro},\ and\ \citenamefont
	{Brown}}]{majumder2020real}%
\BibitemOpen
\bibfield  {author} {\bibinfo {author} {\bibfnamefont {Swarnadeep}\
		\bibnamefont {Majumder}}, \bibinfo {author} {\bibfnamefont
		{Leonardo~Andreta}\ \bibnamefont {de~Castro}}, \ and\ \bibinfo {author}
	{\bibfnamefont {Kenneth~R}\ \bibnamefont {Brown}},\ }\bibfield  {title}
{\enquote {\bibinfo {title} {Real-time calibration with spectator qubits},}\
}\href@noop {} {\bibfield  {journal} {\bibinfo  {journal} {npj Quantum
		Information}\ }\textbf {\bibinfo {volume} {6}},\ \bibinfo {pages} {1--9}
(\bibinfo {year} {2020})}\BibitemShut {NoStop}%
\bibitem [{\citenamefont {Shankar}\ \emph {et~al.}(2013)\citenamefont
	{Shankar}, \citenamefont {Hatridge}, \citenamefont {Leghtas}, \citenamefont
	{Sliwa}, \citenamefont {Narla}, \citenamefont {Vool}, \citenamefont {Girvin},
	\citenamefont {Frunzio}, \citenamefont {Mirrahimi},\ and\ \citenamefont
	{Devoret}}]{shankar2013autonomously}%
\BibitemOpen
\bibfield  {author} {\bibinfo {author} {\bibfnamefont {Shyam}\ \bibnamefont
		{Shankar}}, \bibinfo {author} {\bibfnamefont {Michael}\ \bibnamefont
		{Hatridge}}, \bibinfo {author} {\bibfnamefont {Zaki}\ \bibnamefont
		{Leghtas}}, \bibinfo {author} {\bibfnamefont {KM}~\bibnamefont {Sliwa}},
	\bibinfo {author} {\bibfnamefont {Aniruth}\ \bibnamefont {Narla}}, \bibinfo
	{author} {\bibfnamefont {Uri}\ \bibnamefont {Vool}}, \bibinfo {author}
	{\bibfnamefont {Steven~M}\ \bibnamefont {Girvin}}, \bibinfo {author}
	{\bibfnamefont {Luigi}\ \bibnamefont {Frunzio}}, \bibinfo {author}
	{\bibfnamefont {Mazyar}\ \bibnamefont {Mirrahimi}}, \ and\ \bibinfo {author}
	{\bibfnamefont {Michel~H}\ \bibnamefont {Devoret}},\ }\bibfield  {title}
{\enquote {\bibinfo {title} {Autonomously stabilized entanglement between two
			superconducting quantum bits},}\ }\href@noop {} {\bibfield  {journal}
	{\bibinfo  {journal} {Nature}\ }\textbf {\bibinfo {volume} {504}},\ \bibinfo
	{pages} {419--422} (\bibinfo {year} {2013})}\BibitemShut {NoStop}%
\bibitem [{\citenamefont {Landon-Cardinal}\ \emph {et~al.}(2018)\citenamefont
	{Landon-Cardinal}, \citenamefont {Govia},\ and\ \citenamefont
	{Clerk}}]{govia2018tomography}%
\BibitemOpen
\bibfield  {author} {\bibinfo {author} {\bibfnamefont {Olivier}\ \bibnamefont
		{Landon-Cardinal}}, \bibinfo {author} {\bibfnamefont {Luke C.~G.}\
		\bibnamefont {Govia}}, \ and\ \bibinfo {author} {\bibfnamefont {Aashish~A.}\
		\bibnamefont {Clerk}},\ }\bibfield  {title} {\enquote {\bibinfo {title}
		{Quantitative tomography for continuous variable quantum systems},}\ }\href
{\doibase 10.1103/PhysRevLett.120.090501} {\bibfield  {journal} {\bibinfo
		{journal} {Phys. Rev. Lett.}\ }\textbf {\bibinfo {volume} {120}},\ \bibinfo
	{pages} {090501} (\bibinfo {year} {2018})}\BibitemShut {NoStop}%
\bibitem [{\citenamefont {Caliari}\ \emph {et~al.}(2005)\citenamefont
	{Caliari}, \citenamefont {De~Marchi},\ and\ \citenamefont
	{Vianello}}]{caliari2005bivariate}%
\BibitemOpen
\bibfield  {author} {\bibinfo {author} {\bibfnamefont {Marco}\ \bibnamefont
		{Caliari}}, \bibinfo {author} {\bibfnamefont {Stefano}\ \bibnamefont
		{De~Marchi}}, \ and\ \bibinfo {author} {\bibfnamefont {Marco}\ \bibnamefont
		{Vianello}},\ }\bibfield  {title} {\enquote {\bibinfo {title} {Bivariate
			polynomial interpolation on the square at new nodal sets},}\ }\href@noop {}
{\bibfield  {journal} {\bibinfo  {journal} {Applied Mathematics and
			Computation}\ }\textbf {\bibinfo {volume} {165}},\ \bibinfo {pages}
	{261--274} (\bibinfo {year} {2005})}\BibitemShut {NoStop}%
\bibitem [{\citenamefont {Bos}\ \emph {et~al.}(2006)\citenamefont {Bos},
	\citenamefont {Caliari}, \citenamefont {De~Marchi}, \citenamefont
	{Vianello},\ and\ \citenamefont {Xu}}]{bos2006gencurve}%
\BibitemOpen
\bibfield  {author} {\bibinfo {author} {\bibfnamefont {Len}\ \bibnamefont
		{Bos}}, \bibinfo {author} {\bibfnamefont {Marco}\ \bibnamefont {Caliari}},
	\bibinfo {author} {\bibfnamefont {Stefano}\ \bibnamefont {De~Marchi}},
	\bibinfo {author} {\bibfnamefont {Marco}\ \bibnamefont {Vianello}}, \ and\
	\bibinfo {author} {\bibfnamefont {Yuan}\ \bibnamefont {Xu}},\ }\bibfield
{title} {\enquote {\bibinfo {title} {Bivariate {Lagrange} interpolation at
			the {Padua} points: the generating curve approach},}\ }\href@noop {}
{\bibfield  {journal} {\bibinfo  {journal} {Journal of Approximation Theory}\
	}\textbf {\bibinfo {volume} {143}},\ \bibinfo {pages} {15--25} (\bibinfo
	{year} {2006})}\BibitemShut {NoStop}%
\bibitem [{\citenamefont {Bos}\ \emph {et~al.}(2007)\citenamefont {Bos},
	\citenamefont {De~Marchi}, \citenamefont {Vianello},\ and\ \citenamefont
	{Xu}}]{bos2007ideal}%
\BibitemOpen
\bibfield  {author} {\bibinfo {author} {\bibfnamefont {Len}\ \bibnamefont
		{Bos}}, \bibinfo {author} {\bibfnamefont {Stefano}\ \bibnamefont
		{De~Marchi}}, \bibinfo {author} {\bibfnamefont {Marco}\ \bibnamefont
		{Vianello}}, \ and\ \bibinfo {author} {\bibfnamefont {Yuan}\ \bibnamefont
		{Xu}},\ }\bibfield  {title} {\enquote {\bibinfo {title} {Bivariate {Lagrange}
			interpolation at the {Padua} points: the ideal theory approach},}\
}\href@noop {} {\bibfield  {journal} {\bibinfo  {journal} {Numerische
		Mathematik}\ }\textbf {\bibinfo {volume} {108}},\ \bibinfo {pages} {43--57}
(\bibinfo {year} {2007})}\BibitemShut {NoStop}%
\bibitem [{\citenamefont {Caliari}\ \emph {et~al.}(2011)\citenamefont
	{Caliari}, \citenamefont {De~Marchi}, \citenamefont {Sommariva},\ and\
	\citenamefont {Vianello}}]{caliari2011padua2dm}%
\BibitemOpen
\bibfield  {author} {\bibinfo {author} {\bibfnamefont {Marco}\ \bibnamefont
		{Caliari}}, \bibinfo {author} {\bibfnamefont {Stefano}\ \bibnamefont
		{De~Marchi}}, \bibinfo {author} {\bibfnamefont {Alvise}\ \bibnamefont
		{Sommariva}}, \ and\ \bibinfo {author} {\bibfnamefont {Marco}\ \bibnamefont
		{Vianello}},\ }\bibfield  {title} {\enquote {\bibinfo {title} {{Padua2DM}:
			fast interpolation and cubature at the {Padua} points in
			{Matlab}/{Octave}},}\ }\href@noop {} {\bibfield  {journal} {\bibinfo
		{journal} {Numerical Algorithms}\ }\textbf {\bibinfo {volume} {56}},\
	\bibinfo {pages} {45--60} (\bibinfo {year} {2011})}\BibitemShut {NoStop}%
\bibitem [{\citenamefont {Caliari}\ \emph {et~al.}(2008)\citenamefont
	{Caliari}, \citenamefont {De~Marchi},\ and\ \citenamefont
	{Vianello}}]{caliari2008bivariate}%
\BibitemOpen
\bibfield  {author} {\bibinfo {author} {\bibfnamefont {Marco}\ \bibnamefont
		{Caliari}}, \bibinfo {author} {\bibfnamefont {Stefano}\ \bibnamefont
		{De~Marchi}}, \ and\ \bibinfo {author} {\bibfnamefont {Marco}\ \bibnamefont
		{Vianello}},\ }\bibfield  {title} {\enquote {\bibinfo {title} {Bivariate
			{Lagrange} interpolation at the {Padua} points: Computational aspects},}\
}\href@noop {} {\bibfield  {journal} {\bibinfo  {journal} {Journal of
		Computational and Applied Mathematics}\ }\textbf {\bibinfo {volume} {221}},\
\bibinfo {pages} {284--292} (\bibinfo {year} {2008})}\BibitemShut {NoStop}%
\bibitem [{\citenamefont {Gupta}\ and\ \citenamefont
	{Biercuk}(2019)}]{gupta2019convergence}%
\BibitemOpen
\bibfield  {author} {\bibinfo {author} {\bibfnamefont {Riddhi~S.}\
		\bibnamefont {Gupta}}\ and\ \bibinfo {author} {\bibfnamefont {Michael~J.}\
		\bibnamefont {Biercuk}},\ }\href@noop {} {\enquote {\bibinfo {title}
		{Convergence analysis for autonomous adaptive learning applied to quantum
			architectures},}\ } (\bibinfo {year} {2019}),\ \Eprint
{http://arxiv.org/abs/1911.05752} {arXiv:1911.05752 [quant-ph]} \BibitemShut
{NoStop}%
\bibitem [{\citenamefont {Gupta}\ \emph {et~al.}(2019)\citenamefont {Gupta},
	\citenamefont {Milne}, \citenamefont {Edmunds}, \citenamefont {Hempel},\ and\
	\citenamefont {Biercuk}}]{gupta2019autonomous}%
\BibitemOpen
\bibfield  {author} {\bibinfo {author} {\bibfnamefont {Riddhi~Swaroop}\
		\bibnamefont {Gupta}}, \bibinfo {author} {\bibfnamefont {Alistair~R.}\
		\bibnamefont {Milne}}, \bibinfo {author} {\bibfnamefont {Claire~L.}\
		\bibnamefont {Edmunds}}, \bibinfo {author} {\bibfnamefont {Cornelius}\
		\bibnamefont {Hempel}}, \ and\ \bibinfo {author} {\bibfnamefont {Michael~J.}\
		\bibnamefont {Biercuk}},\ }\href@noop {} {\enquote {\bibinfo {title}
		{Autonomous adaptive noise characterization in quantum computers},}\ }
(\bibinfo {year} {2019}),\ \Eprint {http://arxiv.org/abs/1904.07225}
{arXiv:1904.07225 [quant-ph]} \BibitemShut {NoStop}%
\bibitem [{\citenamefont {Majdisova}\ and\ \citenamefont
	{Skala}(2017)}]{majdisova2017radial}%
\BibitemOpen
\bibfield  {author} {\bibinfo {author} {\bibfnamefont {Zuzana}\ \bibnamefont
		{Majdisova}}\ and\ \bibinfo {author} {\bibfnamefont {Vaclav}\ \bibnamefont
		{Skala}},\ }\bibfield  {title} {\enquote {\bibinfo {title} {Radial basis
			function approximations: comparison and applications},}\ }\href@noop {}
{\bibfield  {journal} {\bibinfo  {journal} {Applied Mathematical Modelling}\
	}\textbf {\bibinfo {volume} {51}},\ \bibinfo {pages} {728--743} (\bibinfo
	{year} {2017})}\BibitemShut {NoStop}%
\bibitem [{\citenamefont {Stein}(2012)}]{stein2012interpolation}%
\BibitemOpen
\bibfield  {author} {\bibinfo {author} {\bibfnamefont {Michael~L}\
		\bibnamefont {Stein}},\ }\href@noop {} {\emph {\bibinfo {title}
		{Interpolation of spatial data: some theory for kriging}}}\ (\bibinfo
{publisher} {Springer Science \& Business Media},\ \bibinfo {year}
{2012})\BibitemShut {NoStop}%
\bibitem [{\citenamefont {Franke}(1979)}]{franke1979critical}%
\BibitemOpen
\bibfield  {author} {\bibinfo {author} {\bibfnamefont {Richard}\ \bibnamefont
		{Franke}},\ }\href@noop {} {\emph {\bibinfo {title} {A critical comparison of
			some methods for interpolation of scattered data}}},\ \bibinfo {type} {Tech.
	Rep.}\ (\bibinfo  {institution} {Naval Postgraduate School Monterey CA},\
\bibinfo {year} {1979})\BibitemShut {NoStop}%
\bibitem [{\citenamefont {Landau}\ \emph {et~al.}(2011)\citenamefont {Landau},
	\citenamefont {Lozano}, \citenamefont {M'Saad},\ and\ \citenamefont
	{Karimi}}]{landau2011adaptive}%
\BibitemOpen
\bibfield  {author} {\bibinfo {author} {\bibfnamefont {Ioan~Dor{\'e}}\
		\bibnamefont {Landau}}, \bibinfo {author} {\bibfnamefont {Rogelio}\
		\bibnamefont {Lozano}}, \bibinfo {author} {\bibfnamefont {Mohammed}\
		\bibnamefont {M'Saad}}, \ and\ \bibinfo {author} {\bibfnamefont {Alireza}\
		\bibnamefont {Karimi}},\ }\href@noop {} {\emph {\bibinfo {title} {Adaptive
			control: algorithms, analysis and applications}}}\ (\bibinfo  {publisher}
{Springer Science \& Business Media},\ \bibinfo {year} {2011})\BibitemShut
{NoStop}%
\bibitem [{\citenamefont {Thrun}\ \emph {et~al.}(2000)\citenamefont {Thrun},
	\citenamefont {Burgard},\ and\ \citenamefont {Fox}}]{thrun2000probabilistic}%
\BibitemOpen
\bibfield  {author} {\bibinfo {author} {\bibfnamefont {Sebastian}\
		\bibnamefont {Thrun}}, \bibinfo {author} {\bibfnamefont {Wolfram}\
		\bibnamefont {Burgard}}, \ and\ \bibinfo {author} {\bibfnamefont {Dieter}\
		\bibnamefont {Fox}},\ }\href@noop {} {\emph {\bibinfo {title} {Probabilistic
			robotics}}},\ Vol.~\bibinfo {volume} {1}\ (\bibinfo  {publisher} {MIT press
	Cambridge},\ \bibinfo {year} {2000})\BibitemShut {NoStop}%
\bibitem [{\citenamefont {Stanley}(2019)}]{wavesense}%
\BibitemOpen
\bibfield  {author} {\bibinfo {author} {\bibfnamefont {B.}~\bibnamefont
		{Stanley}},\ }\href@noop {} {}\bibinfo {howpublished} {personal
	communication, in-person meeting with WaveSense CTO at Greentown Labs,
	Somerville, MA} (\bibinfo {year} {2019})\BibitemShut {NoStop}%
\bibitem [{\citenamefont {Michalek}\ \emph {et~al.}(2000)\citenamefont
	{Michalek}, \citenamefont {Wagner},\ and\ \citenamefont
	{Timmer}}]{michalek2000new}%
\BibitemOpen
\bibfield  {author} {\bibinfo {author} {\bibfnamefont {Steffen}\ \bibnamefont
		{Michalek}}, \bibinfo {author} {\bibfnamefont {Mirko}\ \bibnamefont
		{Wagner}}, \ and\ \bibinfo {author} {\bibfnamefont {Jens}\ \bibnamefont
		{Timmer}},\ }\bibfield  {title} {\enquote {\bibinfo {title} {A new
			approximate likelihood estimator for {ARMA}-filtered hidden {M}arkov
			models},}\ }\href@noop {} {\bibfield  {journal} {\bibinfo  {journal} {IEEE
			Transactions on Signal Processing}\ }\textbf {\bibinfo {volume} {48}},\
	\bibinfo {pages} {1537--1547} (\bibinfo {year} {2000})}\BibitemShut {NoStop}%
\bibitem [{\citenamefont {Rasmussen}\ and\ \citenamefont
	{Nickisch}(2010)}]{rasmussen2010gaussian}%
\BibitemOpen
\bibfield  {author} {\bibinfo {author} {\bibfnamefont {Carl~Edward}\
		\bibnamefont {Rasmussen}}\ and\ \bibinfo {author} {\bibfnamefont {Hannes}\
		\bibnamefont {Nickisch}},\ }\bibfield  {title} {\enquote {\bibinfo {title}
		{Gaussian processes for machine learning ({GPML}) toolbox},}\ }\href@noop {}
{\bibfield  {journal} {\bibinfo  {journal} {Journal of machine learning
			research}\ }\textbf {\bibinfo {volume} {11}},\ \bibinfo {pages} {3011--3015}
	(\bibinfo {year} {2010})}\BibitemShut {NoStop}%
\bibitem [{\citenamefont {Luttinen}\ and\ \citenamefont
	{Ilin}(2012)}]{luttinen2012efficient}%
\BibitemOpen
\bibfield  {author} {\bibinfo {author} {\bibfnamefont {Jaakko}\ \bibnamefont
		{Luttinen}}\ and\ \bibinfo {author} {\bibfnamefont {Alexander}\ \bibnamefont
		{Ilin}},\ }\bibfield  {title} {\enquote {\bibinfo {title} {Efficient gaussian
			process inference for short-scale spatio-temporal modeling},}\ }in\
\href@noop {} {\emph {\bibinfo {booktitle} {Artificial Intelligence and
			Statistics}}}\ (\bibinfo {year} {2012})\ pp.\ \bibinfo {pages}
{741--750}\BibitemShut {NoStop}%
\bibitem [{\citenamefont {Hartikainen}\ \emph {et~al.}(2011)\citenamefont
	{Hartikainen}, \citenamefont {Riihim{\"a}ki},\ and\ \citenamefont
	{S{\"a}rkk{\"a}}}]{hartikainen2011sparse}%
\BibitemOpen
\bibfield  {author} {\bibinfo {author} {\bibfnamefont {Jouni}\ \bibnamefont
		{Hartikainen}}, \bibinfo {author} {\bibfnamefont {Jaakko}\ \bibnamefont
		{Riihim{\"a}ki}}, \ and\ \bibinfo {author} {\bibfnamefont {Simo}\
		\bibnamefont {S{\"a}rkk{\"a}}},\ }\bibfield  {title} {\enquote {\bibinfo
		{title} {Sparse spatio-temporal gaussian processes with general
			likelihoods},}\ }in\ \href@noop {} {\emph {\bibinfo {booktitle}
		{International Conference on Artificial Neural Networks}}}\ (\bibinfo
{organization} {Springer},\ \bibinfo {year} {2011})\ pp.\ \bibinfo {pages}
{193--200}\BibitemShut {NoStop}%
\bibitem [{\citenamefont {Li}\ and\ \citenamefont {Fu}(2012)}]{li2012arma}%
\BibitemOpen
\bibfield  {author} {\bibinfo {author} {\bibfnamefont {Kang}\ \bibnamefont
		{Li}}\ and\ \bibinfo {author} {\bibfnamefont {Yun}\ \bibnamefont {Fu}},\
}\bibfield  {title} {\enquote {\bibinfo {title} {{ARMA-HMM}: A new approach
		for early recognition of human activity},}\ }in\ \href@noop {} {\emph
{\bibinfo {booktitle} {Proceedings of the 21st International Conference on
		Pattern Recognition (ICPR2012)}}}\ (\bibinfo {organization} {IEEE},\ \bibinfo
{year} {2012})\ pp.\ \bibinfo {pages} {1779--1782}\BibitemShut {NoStop}%
\bibitem [{\citenamefont {Van~Barel}\ and\ \citenamefont
	{Humet}(2015)}]{van2015good}%
\BibitemOpen
\bibfield  {author} {\bibinfo {author} {\bibfnamefont {Marc}\ \bibnamefont
		{Van~Barel}}\ and\ \bibinfo {author} {\bibfnamefont {Matthias}\ \bibnamefont
		{Humet}},\ }\bibfield  {title} {\enquote {\bibinfo {title} {Good point sets
			and corresponding weights for bivariate discrete least squares
			approximation},}\ }\href@noop {} {\bibfield  {journal} {\bibinfo  {journal}
		{Dolomites Research Notes on Approximation}\ }\textbf {\bibinfo {volume}
		{8}},\ \bibinfo {pages} {37--50} (\bibinfo {year} {2015})}\BibitemShut
{NoStop}%
\bibitem [{\citenamefont {Ibrahimoglu}(2016)}]{ibrahimoglu2016lebesgue}%
\BibitemOpen
\bibfield  {author} {\bibinfo {author} {\bibfnamefont {Bayram~Ali}\
		\bibnamefont {Ibrahimoglu}},\ }\bibfield  {title} {\enquote {\bibinfo {title}
		{Lebesgue functions and {Lebesgue} constants in polynomial interpolation},}\
}\href@noop {} {\bibfield  {journal} {\bibinfo  {journal} {Journal of
		Inequalities and Applications}\ }\textbf {\bibinfo {volume} {2016}},\
\bibinfo {pages} {93} (\bibinfo {year} {2016})}\BibitemShut {NoStop}%
\bibitem [{\citenamefont {Kolomoitsev}\ \emph {et~al.}(2018)\citenamefont
	{Kolomoitsev}, \citenamefont {Lomako},\ and\ \citenamefont
	{Prestin}}]{kolomoitsev2018lp}%
\BibitemOpen
\bibfield  {author} {\bibinfo {author} {\bibfnamefont {Yurii}\ \bibnamefont
		{Kolomoitsev}}, \bibinfo {author} {\bibfnamefont {Tetiana}\ \bibnamefont
		{Lomako}}, \ and\ \bibinfo {author} {\bibfnamefont {J{\"u}rgen}\ \bibnamefont
		{Prestin}},\ }\bibfield  {title} {\enquote {\bibinfo {title} {On {Lp}-error
			of bivariate polynomial interpolation on the square},}\ }\href@noop {}
{\bibfield  {journal} {\bibinfo  {journal} {Journal of Approximation Theory}\
	}\textbf {\bibinfo {volume} {229}},\ \bibinfo {pages} {13--35} (\bibinfo
	{year} {2018})}\BibitemShut {NoStop}%
\bibitem [{\citenamefont {Stewart}(1996)}]{stewart1996afternotes}%
\BibitemOpen
\bibfield  {author} {\bibinfo {author} {\bibfnamefont {Gilbert~W}\
		\bibnamefont {Stewart}},\ }\href@noop {} {\emph {\bibinfo {title} {Afternotes
			on numerical analysis}}},\ Vol.~\bibinfo {volume} {49}\ (\bibinfo
{publisher} {Siam},\ \bibinfo {year} {1996})\BibitemShut {NoStop}%
\bibitem [{\citenamefont {Austin}\ and\ \citenamefont
	{Trefethen}(2017)}]{austin2017trigonometric}%
\BibitemOpen
\bibfield  {author} {\bibinfo {author} {\bibfnamefont {Anthony~P}\
		\bibnamefont {Austin}}\ and\ \bibinfo {author} {\bibfnamefont {Lloyd~N}\
		\bibnamefont {Trefethen}},\ }\bibfield  {title} {\enquote {\bibinfo {title}
		{Trigonometric interpolation and quadrature in perturbed points},}\
}\href@noop {} {\bibfield  {journal} {\bibinfo  {journal} {SIAM Journal on
		Numerical Analysis}\ }\textbf {\bibinfo {volume} {55}},\ \bibinfo {pages}
{2113--2122} (\bibinfo {year} {2017})}\BibitemShut {NoStop}%
\bibitem [{\citenamefont {Tian}(2005)}]{tian2005moore}%
\BibitemOpen
\bibfield  {author} {\bibinfo {author} {\bibfnamefont {Yongge}\ \bibnamefont
		{Tian}},\ }\bibfield  {title} {\enquote {\bibinfo {title} {The
			{Moore-Penrose} inverse for sums of matrices under rank additivity
			conditions},}\ }\href@noop {} {\bibfield  {journal} {\bibinfo  {journal}
		{Linear and Multilinear Algebra}\ }\textbf {\bibinfo {volume} {53}},\
	\bibinfo {pages} {45--65} (\bibinfo {year} {2005})}\BibitemShut {NoStop}%
\end{thebibliography}


%

\end{document}